\documentclass[aps,pra,twocolumn,groupedaddress]{revtex4} 
\usepackage{epsf}
\usepackage[dvips]{color}
\usepackage{graphicx} 

\begin{document}
\title{Quantized Vortex States of Strongly Interacting Bosons in a
  Rotating Optical Lattice}

\author{Rajiv Bhat$^1$, B.~M.~Peden$^1$, B.~T.~Seaman$^1$, M.~Kr\"amer$^1$, L.~D.~Carr$^2$ and M.~J.~Holland$^1$}

\address{ $^1$JILA and Department of Physics, University of
  Colorado at Boulder, CO 80309-0440, USA \\
  $^2$Physics Department, Colorado School of Mines, Golden, Colorado
  80401, USA}

\begin{abstract}
  Bose gases in rotating optical lattices combine two important topics
  in quantum physics: superfluid rotation and strong correlations.  In
  this paper, we examine square two-dimensional systems at zero
  temperature comprised of strongly repulsive bosons with filling
  factors of less than one atom per lattice site. The entry of
  vortices into the system is characterized by jumps of $2\pi$ in the
  phase winding of the condensate wavefunction.  A lattice of size
  $L\times L$ can have at most $L-1$ quantized vortices in the lowest
  Bloch band. In contrast to homogeneous systems, angular momentum is
  not a good quantum number since the continuous rotational symmetry
  is broken by the lattice. Instead, a quasi-angular momentum captures
  the discrete rotational symmetry of the system. Energy level
  crossings indicative of quantum phase transitions are observed when
  the quasi-angular momentum of the ground-state changes.
  
\end{abstract}
\maketitle

\section{Introduction}

Rotating Bose-Einstein condensates in dilute alkali gases provide a
rich playground for the study of quantized vortices in superfluid
systems. One major advantage is the ability to directly image vortex
cores and study their static and dynamic properties. Vortices, first
made by quantum state engineering of condensate wave functions
~\cite{Matthews:1999,Williams:1999}, are typically now produced by mechanical
stirring of ultracold atomic clouds ~\cite{Madison:2000,Haljan:2001}.
Remarkable images of large vortex lattices containing more than one
hundred vortices in an Abrikosov type triangular configuration have
produced striking evidence for the superfluidity of Bose-Einstein
condensed alkali gases~\cite{Aboshaeer:2001,Coddington:2003}.

Employing the analogy between the Hamiltonian for a two-dimensional
electron gas in a strong magnetic field and that for a rotating
atomic gas, it has been pointed out that the physics of the fractional
quantum Hall effect (FQHE) should emerge when the number of vortices
and the number of atoms become comparable
~\cite{Wilkin:2000,Paredes:2001,Fischer:2004,Baranov:2004}. Achieving
this regime experimentally is a significant goal of the field at this
time. One of the reasons for this is the connection with many problems
of interest in condensed matter systems where strongly correlated
electron effects have been discussed and studied. The direct approach
of spinning-up a Bose-Einstein condensate to reach the regime of
strongly correlated effects is difficult because of the need to reach
a parameter regime of low particle number per vortex and extremely low
temperature~\cite{Schweikhard:2004}; for this reason the FQHE regime
has yet to be achieved with cold atoms.

Ultracold gases are typically dilute, and the interaction effects can
be well incorporated by perturbation theory. However, it is possible
to manipulate and enhance the interaction effects in a number of ways
so that the perturbative treatment fails.  One possibility is to
increase the two-body scattering length via a Feshbach resonance.  An
alternative method is to modify the effective interactions through
application of an optical lattice. An optical lattice is formed from
an off-resonant light intensity pattern created by the interference of
several laser beams.  The atoms feel a potential proportional to the
intensity of the light field. As the laser fields are made more
intense, the interactions are enhanced because the atoms become more
strongly confined in the lattice wells. This technique was exploited
in the theoretical and experimental study of the Mott-insulator to
superfluid quantum phase transition in a non-rotating system
~\cite{Jaksch:1998,Greiner:2002}.  Theoretical studies have been done linking the
Hofstadter butterfly ~\cite{Jaksch:2003} and the FQHE
~\cite{Palmer:2006, Soerenson:2005} with bosons in an optical lattice in the presence
of an effective magnetic field. 

It is thus a natural question to pose as to whether one can combine
the intriguing physics of the rotating gas with the enhanced
interactions in an optical lattice. The motivation would be to move
the regime of quantum Hall physics towards a parameter space that is
experimentally achievable. In a first experiment with a rotating
optical lattice, Tung {\it et al.}~\cite{Tung:2006} have recently
demonstrated vortex pinning in a weakly interacting BEC. This was realized
by passing a laser beam through a mask that contains holes arranged in
a particular configuration and then focusing the laser beams to form
the lattice interference pattern. The two dimensional optical lattice is rotated by spinning the mask.

In this paper, we formulate a theoretical description of
bosons in rotating optical lattices and illuminate the connections
with vortex physics, building upon results presented in an earlier
article~\cite{Bhat:2006}. Certain aspects of this problem such as
single vortex formation~\cite{Wu:2004} and vortex
pinning~\cite{Reijnders:2005} have been theoretically explored for
high filling factors (large number of atoms per site). Burkov {\it
  et.al.}~\cite{Burkov:2006} have shown the formation of delocalized
vortex clusters in lattices containing superfluids in the presence of
an {\em effective} magnetic field created by modulating the optical
lattice. We consider strongly repulsive bosons
in a small 2D rotating square optical lattice with filling factors
less than unity. In this regime, exact solutions are tractable and
indicate key properties of larger systems.

The system can be studied using a modified Bose-Hubbard Hamiltonian
with a complex, site-dependent hopping term which is sufficient to
describe physics in the lowest Bloch band.  The lattice breaks the
continuous rotational symmetry associated with the angular momentum operator. Instead, it possesses a discrete rotational symmetry, and the
generator of this discrete rotation, the quasi-angular momentum, plays
an important role. The square lattice is four-fold rotationally symmetric and 
the quasi-angular momentum operator generates rotations in steps of $\pi/2$. 

The entry of vortices into the system is marked by $2\pi$ jumps in the
phase winding of the condensate wavefunction. We show that a maximum
vorticity of $2\pi(L-1)$ is possible for the lowest band in a lattice
of size $L\times L$. Changes in the quasi-angular momentum of the
ground state are associated with energy level-crossings as a function
of changing angular velocity. For filling commensurate with the
symmetry of the system, the quasi-angular momentum is zero at all
angular velocities. In this case, avoided energy level crossings of the
ground-state are observed as a function of lattice angular velocity. For
incommensurate filling, there are energy level crossings for many
particles at zero temperature. Since these correspond to a symmetry
change in the ground state as a function of a Hamiltonian parameter at
zero temperature, and the property holds for systems of arbitrary size
and particle number, these critical points are {\em quantum phase transitions}
~\cite{Sachdev:2001}.

This paper is structured as follows. Section~\ref{Section:Hamiltonian}
provides the theoretical framework by sketching the derivation of a
modified Bose-Hubbard Hamiltonian for describing lowest-band physics
and discusses its regime of validity, along with the principal
observables. Section~\ref{Section:General} describes general
characteristics of the system such as the Mott-insulator/superfluid
phase diagram and the effect of strong repulsive interactions. The
behavior of the density distribution as a function of rotation is
illustrated. Section~\ref{Section:QAM} presents a characterization of
the quasi-angular momentum.  Sections~\ref{Section:SingleParticle} and
\ref{Section:ManyParticles} discuss results for a single particle and
many particles in the system, respectively. Section~\ref{Section:Conclusions} 
summarizes the main results of this paper.

\section{Modified Bose-Hubbard Hamiltonian for rotating lattices
\label{Section:Hamiltonian}}

\begin{figure}[t]
\begin{center}
   \includegraphics[width=6.0cm]{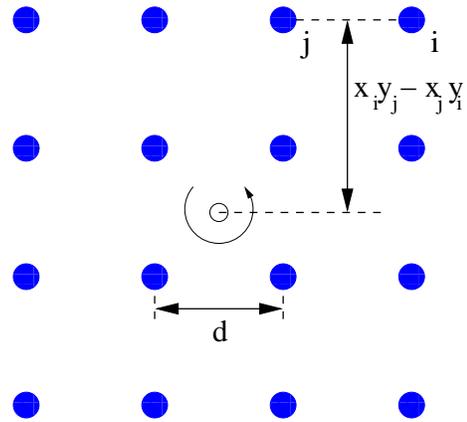}
   \caption{ Schematic for a $4 \times 4$ lattice rotating in the
     counter-clockwise direction. Sites labeled $i$ and $j$ are
     nearest neighbors. The rotation-driven hopping between
     neighbouring sites is governed by the parameter $K _{ij}=\beta (x
     _i y_j -x_j y_i)/d$ where $\beta$ is given by the overlap
     integral Eq.~(\ref{beta}) and the second factor is the
     perpendicular distance of the line joining the two neighboring
     sites from the center of rotation.
     \label{Schematic}}
\end{center}
\end{figure}

\subsection{Hamiltonian}
The energy for a fixed number of bosons in an optical lattice can be
broken down into three components corresponding to the kinetic energy,
the potential energy due to the lattice, and the interaction energy between
bosons.  Using a standard procedure~\cite{Landau:Mechanics}, the
Hamiltonian in the reference frame rotating with angular velocity
$\Omega$ about the $z$-axis is $\hat{H}=\hat{H}_0-\int d\mathbf{x}\hat{\Phi}^{\dagger}\Omega L_z\hat{\Phi}$, where $\hat{H}_0$ is the 
Hamiltonian in the laboratory frame and $L_z$ is the angular momentum. 
This coordinate transformation facilitates the calculation of the ground state in the
laboratory frame since it renders the Hamiltonian time-independent. The
Hamiltonian in the rotating frame reads
\begin{equation}
  \hat{H}=\int  d {\bf x} \hat{\Phi}^{\dagger}\left[  -\frac{\hbar ^2}{2M} 
    \nabla ^2 + V ^{\mathrm{lat}} ({\bf x})+\frac{g}{2} \hat{\Phi}^{\dagger} \hat{\Phi}
    -\Omega L_z\right] \hat{\Phi}, \label{H1}
\end{equation} 
where $M$ is the single particle mass and $g$ is the coupling constant
for repulsive two-body scattering in a dilute gas. The bosonic field operator $\hat{\Phi}$ obeys the commutation relationship
$\left[\hat{\Phi}({\bf x}) ,{\hat\Phi}({\bf x'})^{\dagger}\right]
=\delta({\bf x}-{\bf x'})$.  Particles can be described using an
orthonormal Wannier basis $W^{(p)} _i ({\bf x})$.  Here, $i$ indexes
the $N$ sites on the lattice and $p$ denotes the band index
~\cite{Jaksch:1998}. If the energy due to interaction and rotation is
small compared to the energy separation between the lowest and first
excited band, the particles are confined to the lowest Wannier
orbital.  We shall consider this regime only and henceforth will
drop the band index $p$.  The field operator $\hat{\Phi}$ can be
expanded in terms of this Wannier basis, $W _i ({\bf x})$, and the
corresponding site-specific annihilation operators, $\hat{a} _i$, as
\begin{equation}
  \hat{\Phi}({\bf x})=\sum _{i=1} ^{N} \hat a _i W _i ({\bf x}) \,.\label{phi}
\end{equation}
Alternatively, a rotation dependent phase can be ascribed to each
Wannier basis element and the field operator $\hat{\Phi}$ expanded
accordingly. Comparisons with this approach are made in the next
subsection.

\begin{figure}[t]
\begin{center}
   \includegraphics[width=8.5cm]{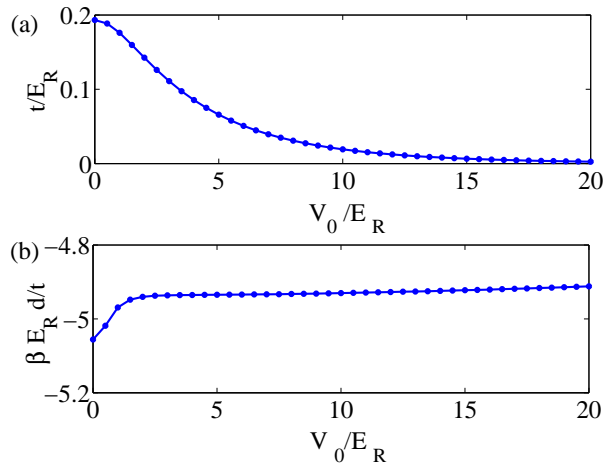}
\caption{ Overlap integrals $t$ (Eq.~(\ref{t})) and $\beta$
   (Eq.~(\ref{beta})) for a standing wave optical lattice described by
   $V=V _0 (\sin ^2 (\pi x/d)+\sin ^2 (\pi y/d))$. The lattice depth
   is given in units of the recoil energy $E_R=\hbar^2\pi^2/2md^2$.
   (a) The hopping parameter $t$ decreases exponentially as a function
   of $V _0$. (b) $\beta dE_R/t$ as a function of lattice depth.  In the 
   tight-binding regime, $\beta d $ scales roughly linearly with $t/E_R$. 
    In this paper we use $\beta =4.93 t/E_Rd$ and $t=0.02E_R$ , corresponding 
    to a standing wave optical lattice of depth $V_0/E _R = 10$.
  \label{Overlap} }
\end{center}
\end{figure}

In the tight binding regime, tunneling between sites which are not
nearest neighbors can be neglected. The interaction between particles 
on nearest neighbor sites can also be neglected.  Using this approximation 
and substituting Eq.~(\ref{phi}) into Eq.~(\ref{H1}) yields the 
modified Bose-Hubbard Hamiltonian
\begin{eqnarray}
  \hat{H}&=&-t \! \sum _{\langle i,j \rangle} \left(  \hat{a} _i ^{\dagger} 
    \hat{a} _j +\hat{a}_i \hat{a} _j ^{\dagger}\right) +\epsilon \sum _i 
  \hat{n} _i + \frac{U}{2}\sum _i \hat{n}_i (\hat{n}_i - 1) \nonumber \\
  &&-i\hbar \Omega \! \sum _{\langle i,j \rangle} \! K _{ij}\left(  
    \hat{a} _i ^{\dagger} \hat{a} _j - \hat{a}_i 
\hat{a} _j ^{\dagger}\right),\label{MBH} 
 \end{eqnarray} 
 where $i$ and $j$ are site indices, $\langle i,j \rangle$
 indicates that the sum is over nearest neighbors, and $\hat{n}_i$ is the
 number operator for site $i$. The first three terms are common to the
 well-studied Bose-Hubbard Hamiltonian for particles in a stationary
 lattice~\cite{Fisher:1989}. The parameters $t$ and
 $\epsilon$ are integrals describing hopping and
 onsite zero-point energy respectively:
\begin{eqnarray}
  t& \equiv &\int d{\bf x} W ^{*} _i ({\bf x}) 
  \left[ -\frac{\hbar ^2}{2M} \nabla ^2 + 
    V ^{(lat)} ({\bf x}) \right] W _j ({\bf x}) \label{t}\,, \\
  \epsilon & \equiv & \int d{\bf x} W ^{*} _i ({\bf x}) 
  \left[ -\frac{\hbar ^2}{2M} \nabla ^2 + 
    V ^{(lat)} ({\bf x}) \right] W _i ({\bf x}) \,.\label{eps} 
\end{eqnarray}   
Wannier functions along the $x$ and $y$ directions can be decoupled
for a square lattice, and accordingly, the integrals
become one-dimensional. The third term in the Hamiltonian, Eq.~(\ref{MBH}), describes the interaction between particles on the
same site. For an $s$-wave scattering length $a_s$
~\cite{Jaksch:1998,Zwerger:2003},
\begin{eqnarray}
  U\equiv\frac{4\pi a_s \hbar ^2}{M} \int d{\bf x}\left|  W_i ({\bf x})\right|^4 .
\end{eqnarray}   
The last term in Eq.~(\ref{MBH}) is the modification due to the rotation
and favors hopping along one azimuthal direction. $K _{ij}$ is a
product of the azimuthal overlap integral, $\beta$, which is dependent
on the geometry and form of the lattice, and the perpendicular distance of the
line joining sites $i$ and $j$ from the center of rotation
(Fig.~\ref{Schematic}),
\begin{eqnarray}
  K _{ij} &=& \frac{\beta}{d} (x _i y_j -x_j y_i) \,.\label{Kij} 
\end{eqnarray}
Here, ($ x _i$,$y _i$) are the coordinates of the $i ^{th}$ site with
the origin located at the center of rotation. The lattice
spacing  is $d$ and 
\begin{eqnarray}
\beta &\equiv& \int d x W _i ^* (x-d) \partial _x W _j  (x) \label{beta}
\end{eqnarray}
We numerically calculate $t$ and $\beta$ for a sinusoidal lattice
potential, $V=V _0 (\sin ^2 (\pi x/d)+\sin ^2 (\pi y/d))$ using
Mathieu functions. The results are plotted in Fig.~\ref{Overlap} as a
function of the lattice depth up to a very tight confinement of
$V_0/E_R=20$, with $E_R=\hbar^2\pi^2/2md^2$ denoting the recoil
energy.

 The eigenstates of the Hamiltonian can be written in the form 
 \begin{equation}
 |\Psi\rangle = \sum _{ \{ n_i \} } c _{ \{ n_i \} } 
|n_1,n_2,\ldots,n_N\rangle \,.\label{prod}
 \end{equation}
 Here $N$ indicates the total number of sites and $\{ n _i \}$
 indicates the set of all possible products of number states
 constrained by the total number of particles, i.e. $\sum_i n_i=n$. In
 this paper, we use the truncated set of Fock states
 $\left\{|0\rangle,|1\rangle\right\}$ to describe the number of
 particles at each site~\cite{Soerenson:2005,Carr:2005}. This corresponds to
 assuming a regime of strong repulsive interactions, i.e., that of hard-core 
 bosons.  Note that in a 1D lattice this approach
 is equivalent to mapping bosonic operators onto fermionic ones via
 the Jordan-Wigner transformation~\cite{Jordan1928,Sachdev:2001}. A
 test of the regime of validity of this approximation is provided in
 Section \ref{Subsection:Effectofinteraction}. The Hamiltonian is
 constructed using this basis and diagonalized to find the ground
 state energy eigenvalue and eigenstate. A set of tools, which are
 described in Section~\ref{Section:Toolkit}, are then used to analyze the ground
 state. Note that solving for the Hamiltonian in Eq.~(\ref{MBH}) is
 equivalent to putting the 2D lattice inside a box with infinite
 potential walls. This effect leads to a number density distribution
 that is peaked in the center for zero rotation.

 \subsection{Regime of Validity of the Hamiltonian}

\begin{figure}[t]
\begin{center}
   \includegraphics[width=8.6cm]{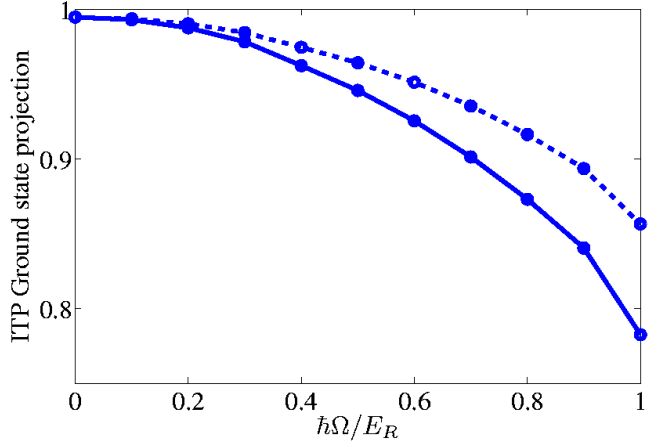}
   \caption{Projection of the ground state
   obtained using imaginary time propagation (ITP) onto the Hilbert
   space spanned by the eigenvectors obtained using $\hat{H}$
   (solid line) (Eq.~(\ref{MBH})) and $\hat{H}_2$ (dashed line) (Eq.~(\ref{H2})) 
   for a $2\times2$ lattice with a lattice depth of $V_0=10E_R$.  $\Omega$ 
   is in units of the recoil energy. The overlap is good even up to
   $\Omega=E_R/\hbar\sim 50t$ where $t$ is the hopping
   energy. \label{ComparisonHilbert}}
\end{center}
\end{figure}

\begin{figure}[t]
\begin{center}
   \includegraphics[width=8.6cm]{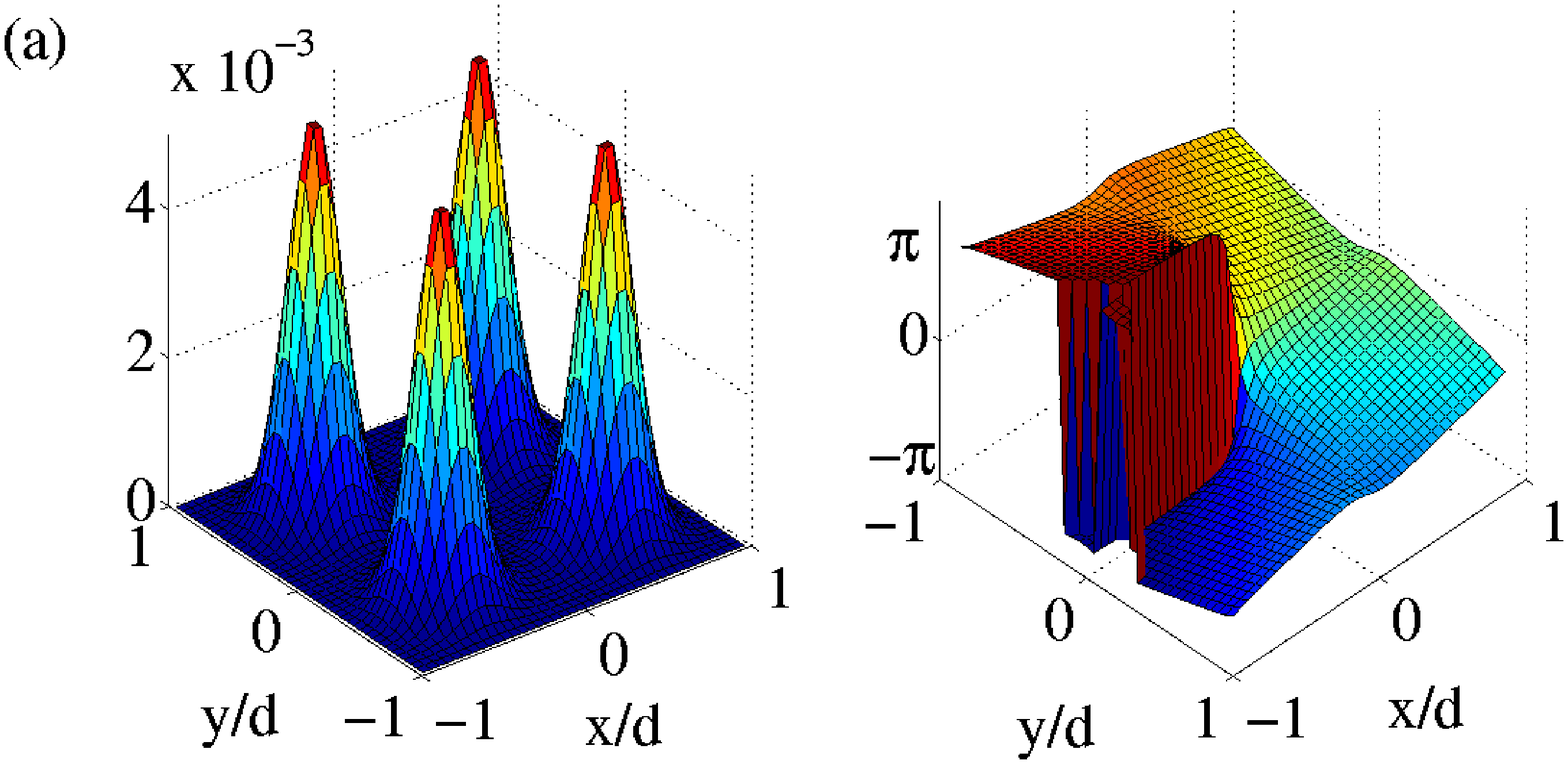}
   \includegraphics[width=8.6cm]{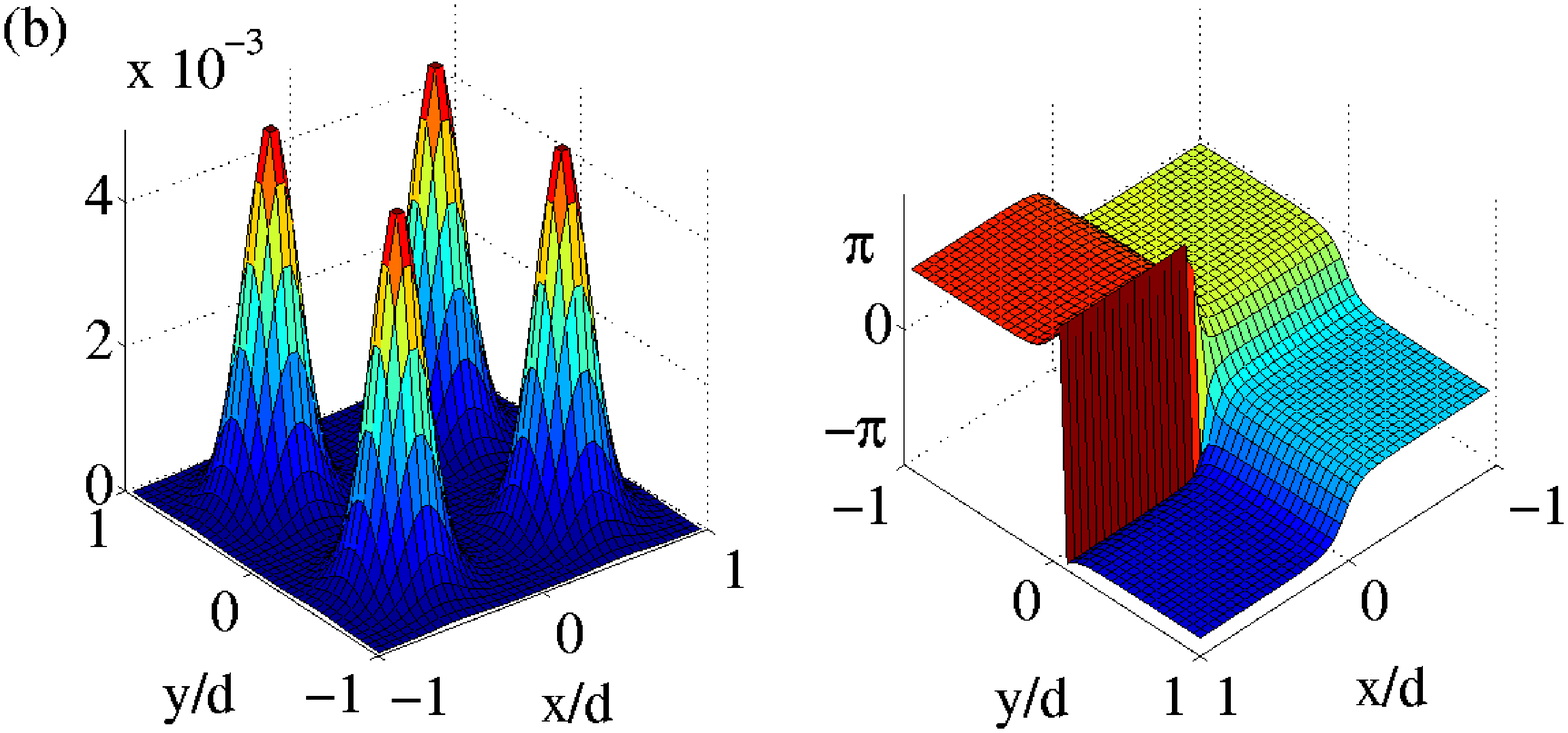}
   \includegraphics[width=8.6cm]{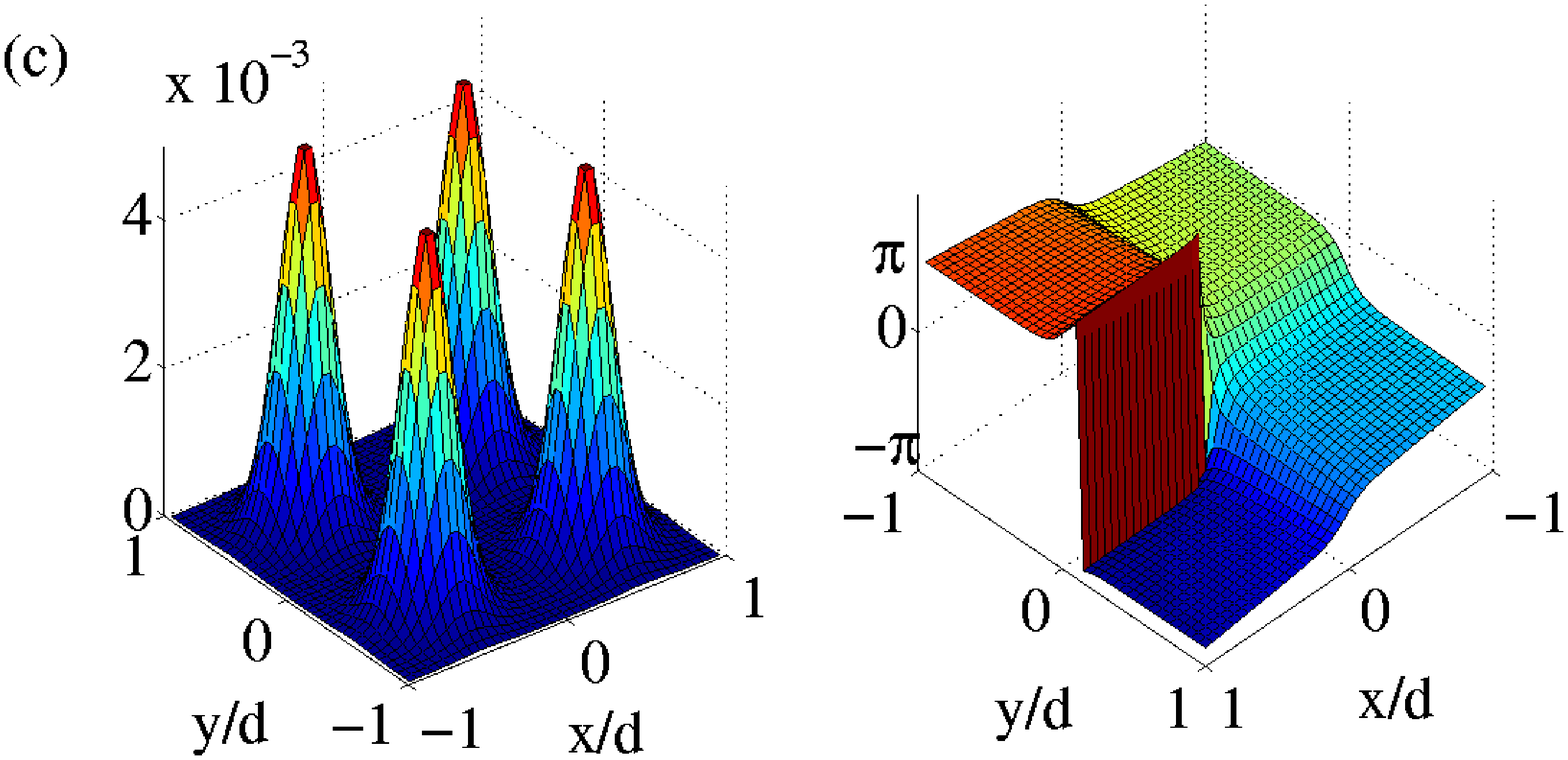}
   \caption{Spatial ground state number density (left) and phase
   information (right) for one particle in a $2\times 2$  sinusoidal 
   lattice with $\Omega=0.5 E_{R}/\hbar$ and
   $V_{0}=10E_{R}$ obtained using (a) Imaginary time propagation, (b)
   Hamiltonian $\hat{H}_{1}$ (Eq.~(\ref{MBH})) and (c) Hamiltonian
   $\hat{H}_{2}$ (Eq.~(\ref{H2})).\label{Comparison}}
\end{center}
\end{figure}

Many aspects of studying the characteristics of bosons in a rotating
optical lattice map onto the extensively studied problem of Bloch
electrons in the presence of a magnetic
field~\cite{Kohn:1959,Blount:1961,Wannier:1962,Hofstadter:1976}. This
subsection makes a connection with the electron problem, while
exploring, in parallel, the limitations of using the lowest band
Bose-Hubbard Hamiltonian.

The Hamiltonian described in Eq.~(\ref{MBH}) only takes into account
Wannier orbitals from the lowest band---an excellent approximation for
small $\Omega$ and a deep lattice. An alternative formulation can be obtained 
by using Eq.~(\ref{H1}) in conjunction with a different Wannier basis given by
 \begin{equation}
   W'_i ({\bf x})=\exp\left[-\frac{iM}{\hbar}\int _{\textbf{x}_i} ^{\textbf{x}}
   \textbf{A} (\textbf{x}')\cdot d\textbf{x}'\right]W _i ({\bf x}), \label{W2}
\end{equation}
resulting, for the case of a single particle in which interaction effects are not
present, in the following Hamiltonian,
\begin{eqnarray}
\hat{H}_2&=&-\sum _{\langle i,j \rangle} \left( t+\frac{1}{2} m\Omega
   ^2 t' \right) \hat{a}^{\dagger} _i \hat{a}_j \nonumber \\
   &&\times \exp\left[-\frac{im}{\hbar}\int _{\textbf{x}_j} ^{\textbf{x}_i} \textbf{A}
   (\textbf{x}')\cdot d\textbf{x}'\right] +h.c. \nonumber \\ 
   & &+\sum _i \left(\epsilon -\frac{1}{2}m\Omega ^2 r_i ^2-\frac{1}{2}
   m\Omega^2\epsilon '\right) \hat{a}^{\dagger} _i \hat{a}_i .
   \label{H2}
\end{eqnarray}
Here, $t'$ and $\epsilon '$ refer to integrals similar to
those defined for $t$ and $\epsilon$ in Eqs.~(\ref{t}) and
(\ref{eps}).  $\textbf{A}(\textbf{x})={\mathbf\Omega}\times\textbf{x}$ is the analog
of the magnetic vector potential. For more than one particle, this
Hamiltonian must be extended to include the effects of interactions.  
Note that $\hat{H} _2$ and $W' _i ({\bf x})$ have forms similar to 
those used traditionally in the treatment of Bloch electrons 
in the presence of magnetic fields. A similar formulation has also been used 
to study bosons in an optical lattice in the presence of an effective magnetic field~\cite{Jaksch:2003,Palmer:2006} and for bosons in a rotating optical lattice~\cite{Wu:2004}.

In this subsection, the two approaches are compared with results for a single 
particle obtained from imaginary time propagation (ITP) for one particle in a $2
\times 2$ lattice.  This analysis gives rise to three findings: (1)
The ground state of the Hamiltonian described in Eq.~(\ref{MBH}) no
longer depends on the increase in $\Omega$ once a maximum phase
difference of $\pi/2$ between neighboring sites has been reached, i.e., 
all the vortex entry transitions possible within the lowest
band have occurred.  For one particle in a $2\times 2$ lattice, the
corresponding maximum phase winding is $2\pi$. This limitation does
not apply to the Hamiltonian Eq.~(\ref{H2}). (2) The Hilbert spaces
spanned by the eigenstates of both Hamiltonians Eq.~(\ref{MBH}) and
Eq.~(\ref{H2}) capture most of the exact ground state wavefunction for
$\hbar\Omega\leq E_R$. Note that $\hbar\Omega\sim E_R$ is large from an experimental point of view. For the case of one particle in a $2\times 2$ lattice, the projection of the exact wavefunction on either Hilbert space is
$\geq 90 \%$ for $\hbar \Omega \sim 0.5 E_R$
(Fig.~\ref{ComparisonHilbert}). Both approaches yield accurate density
profiles for large $\Omega$ ($\sim E_R/\hbar$) but differ from the
ITP-result, and from each other, with regard to the velocity pattern.
Note that $\hat H$ and $\hat H_2$ involve different approximations to
the phase gradient.  The Hamiltonian $\hat H$ allows for phase changes
only in the region of overlap of next-neighbor Wannier functions $W
_i ({\bf x})$, i.e., yields a uniform phase around the site center
(Fig.~\ref{Comparison}(b)). The Hamiltonian $\hat H_2$ requires phase
gradients to be proportional to $\Omega$ and allows for non-zero phase
gradients within each well (Fig.~\ref{Comparison}(c)).  (3) The
lattice rotation frequencies at which the first vortices appear, as will
be discussed later, are slightly different in the two cases over the
range of interest due to the different influence of higher bands in
the three formulations.  The remainder of our work is primarily
concerned with the states of the system at low $\Omega$ and since 
these are well captured by the simpler lowest band Hamiltonian 
described in Eq.~(\ref{MBH}), we use it for the rest of the paper.

\subsection{Toolkit \label{Section:Toolkit}}

We evaluate  six quantities to characterize the behavior of the
system: (1) energy, (2) site number density, (3) intersite current in
the rotating frame,  (4) average angular momentum, (5)
quasi-angular momentum eigenvalues and (6) phase winding of the
condensate wavefunction. 

The ground state energy is obtained as the lowest eigenvalue of the
Hamiltonian. The site number density is the expectation value of the
number operator: $n _i =\langle \hat{a}^{\dagger} _i \hat{a} _i
\rangle$.  The expectation value for the current $J _{ij}$ flowing
from site $i$ to a neighboring site $j$ in the rotating frame is
obtained using the continuity equation,
\begin{eqnarray}
J _{ij}&=&-\frac{1}{i\hbar} \langle [\hat{n}_i,\hat{H}_{ij} ]
\rangle \nonumber \\
 &=&\frac{it}{\hbar}\langle \hat{a}_i \hat{a}_j ^{\dagger}-
\hat{a}_i ^{\dagger} \hat{a}_j\rangle -\Omega K _{ij} \langle 
\hat{a}_i \hat{a}_j ^{\dagger}+\hat{a}_i ^{\dagger}
\hat{a} _j\rangle, \label{Jij}
\end{eqnarray} 
where the current is in units of $ t/\hbar$. 
$\hat{H} _{ij}$ in Eq.~(\ref{Jij}) is the part of the Hamiltonian
operator relevant to sites $i$ and $j$. Since the number density on
any site $i$ is constant for any steady state solution, the algebraic
sum of currents associated with any site $i$ is zero.

The derivative of the energy with respect to the angular velocity
$\Omega$ --- keeping all other Hamiltonian parameters constant --- gives
direct access to the average angular momentum,
\begin{equation}
  \langle {\hat L}_z \rangle = -\frac{\partial E}{\partial\Omega}. \label{L}
\end{equation}
Quasi-angular momentum is calculated based on the four-fold rotational
symmetry in the Hamiltonian created by the square lattice, i.e., 
rotating the system by $90$ degrees has no effect on the
Hamiltonian.  This is discussed in detail in Section
\ref{Section:QAM}.  Here, we give a brief description of the way we
calculate the quasi-angular momentum.  A discrete rotational symmetry
operator, ${\bf R}(\pi/2)$, can be constructed such that ${\bf
  R}(\pi/2)$ acting on a wavefunction rotates that wavefunction by
ninety degrees. ${\bf R}(\pi/2)$ commutes with the Hamiltonian and
therefore shares simultaneous eigenfunctions. Applying this operator
four times corresponds to a rotation of $2\pi$, bringing the
wavefunction back to its original state:
\begin{eqnarray}
{\bf R}\Psi &=& r \Psi \nonumber\\
{\bf R}^4 \Psi &=& \Psi \Rightarrow r=e ^{im\pi/2}, 
\quad m \in \{0,1,2,3\} \,.\label{m}
\end{eqnarray}
Hence, the eigenvalues of $\bf{R}(\pi /2)$ are defined by the
quantized dimensionless quasi-angular momentum $m$. From an operational 
standpoint, we apply a rotation $\bf{R}(\pi/2)$ to $\Psi$ and read out the eigenvalue, 
$r$. Note that the discrete rotational symmetry operator can be generalized to 
a $n-$fold rotationally symmetric system.

The condensate wavefunction is the eigenfunction corresponding to a
macroscopically large eigenvalue of the one-body density matrix $G
^{(1)}$, with all other eigenvalues being non-macroscopic
~\cite{Penrose:1956,Pitaevskii:2003}. In usual tensor notation, $G
^{(1)} _{ij}=\langle \hat{a} ^{\dagger} _j \hat{a} _i \rangle$.  The
phase of the condensate wavefunction describes the superfluid
properties of the system.  In the small systems we consider in this
paper, this is not a rigorous definition. However, a meaningful 
condensate wavefunction can still be obtained in this way since 
in the superfluid regime one of the eigenvalues of the one-body 
density is always significantly larger than all the others, even 
for a very small number of particles. The phase winding around 
the perimeter of the condensate wavefunction,
$\Theta _{cf}$, when divided by $2\pi$ gives the vorticity of the
system. The subscript $cf$ indicates that the phase winding refers to
that of the condensate wavefunction.

\section{General characteristics \label{Section:General}}

This section describes four general features of bosons in rotating
lattices: (1) interaction effects, (2) the Mott insulator/superfluid
quantum phase diagram for the system, (3) number density depletion at
the center for odd lattices, and (4) number density distribution for
even lattices. The first subsection lays out the justification for our choice of
studying lattices using a truncated Fock space as mentioned above in
connection with Eq.~(\ref{prod}). The second subsection connects with existing
understanding of the phase diagram for the Bose-Hubbard model.  The 
third subsection discusses our choice of lattices with an even number of sites 
and the last describes number density rearrangement with vortex entry.

\subsection{Effect of interaction\label{Subsection:Effectofinteraction}}

\begin{figure}[t]
\begin{center}
   \includegraphics[width=8.6cm]{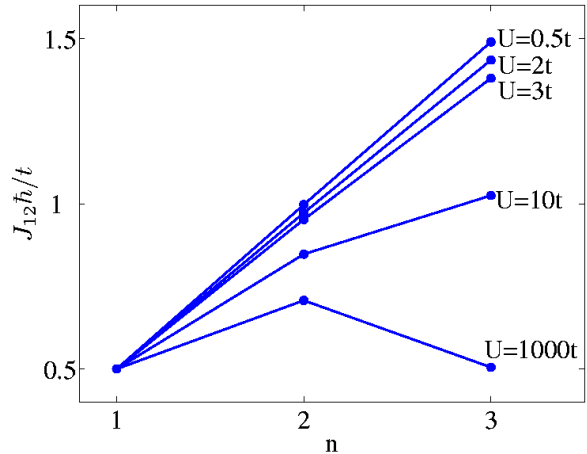}
   \caption{Current $J _{12}$ between adjacent sites of a $2 \times 2$
     unit cell lattice as a function of total number of particles $n$
     for different values of the ratio between repulsive interaction
     and hopping $U/t$ at large rotation ($\hbar \Omega \sim 0.6
     E_R\gg t$).  For weak interaction, $U=0.5t$, the particles can
     freely cross each other and the current is proportional to the
     number of particles. As the interaction increases, the current
     per particle drops as a function of filling. For $U=1000t$, the
     current for three particles is the same as that for one particle,
     indicating a particle-hole symmetry. The current was evaluated
     using a four-state Fock basis on each site, which allowed three 
     particles per site.\label{InteractionCurrent}}
\end{center}
\end{figure}

Interaction between bosons inhibits current flow by making it
difficult for particles to cross each other. To demonstrate this we
consider currents in a $2\times2$ unit lattice using a four-state Fock
basis on each site. Figure~\ref{InteractionCurrent} is a plot of the
current in the rotating frame between two neighboring sites as a function of
filling for different interaction strengths at fixed angular velocity
($\hbar \Omega \sim 0.6 E_R \gg t$). For weak interactions ($U=0.5t$),
the current is proportional to the number of particles as they can
flow independently of each other. However, the current per particle drops with
increasing interaction and filling. At large interaction ($U=1000t$), the current
for three particles (one hole) is the same as for one particle in the
system.  This particle-hole symmetry is characteristic of the regime
where bosons are impenetrable, i.e., of the regime where the
two-state approximation applies~\cite{Soerenson:2005,Carr:2005}. In fact, 
currents calculated for $U\ge 100t$ using the two-state approximation 
coincide with those obtained with a larger Fock space.  The main results 
of this paper are obtained assuming the atoms to be impenetrable and 
hence are expected to be quantitatively accurate in the regime 
$U\ge 100t$ for fillings $\le 1$.

\subsection{Phase diagram}

\begin{figure}[t]
\begin{center}
   \includegraphics[width=8cm]{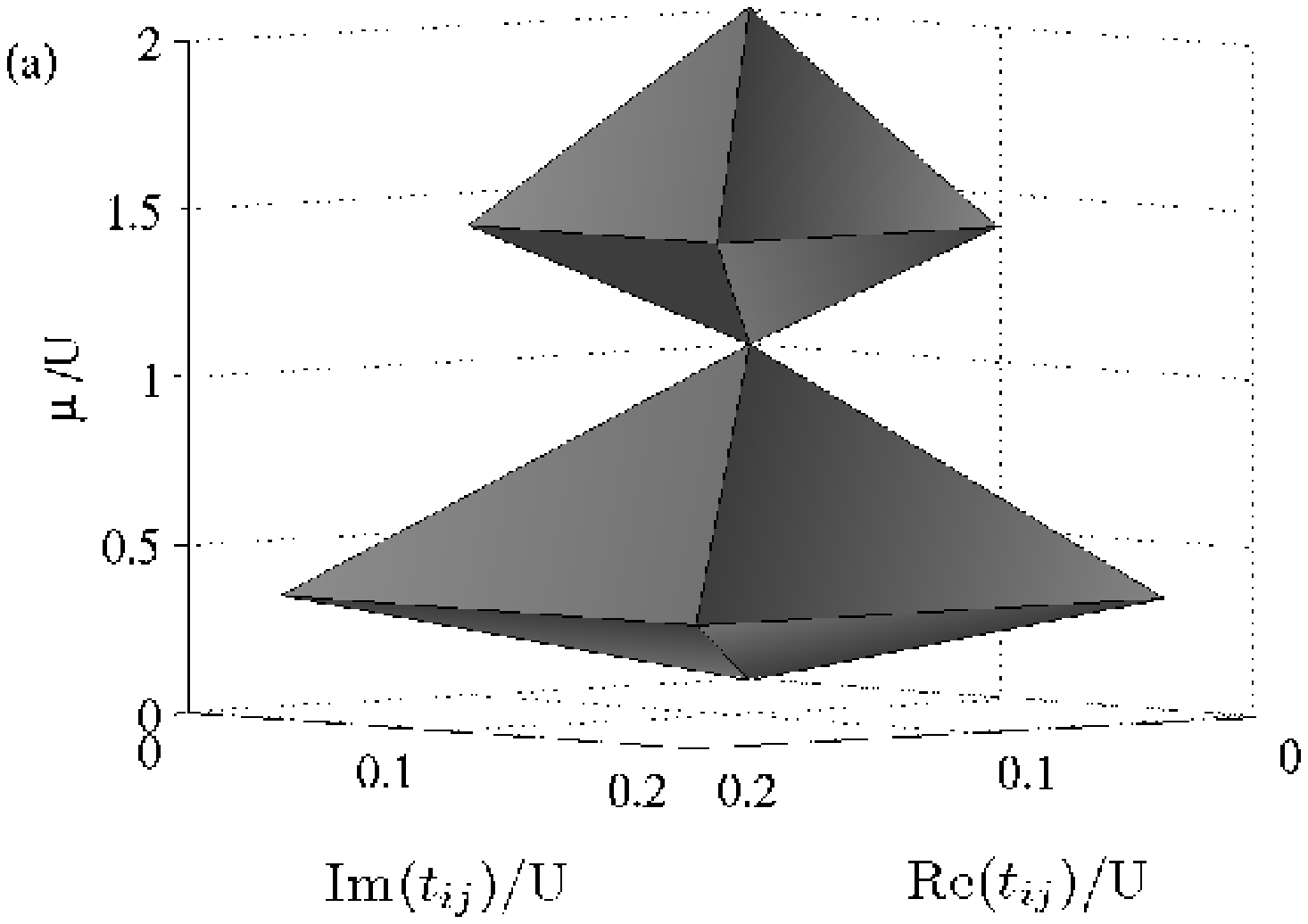}
    \begin{minipage}{8.6 cm}
         \includegraphics[width=4cm]{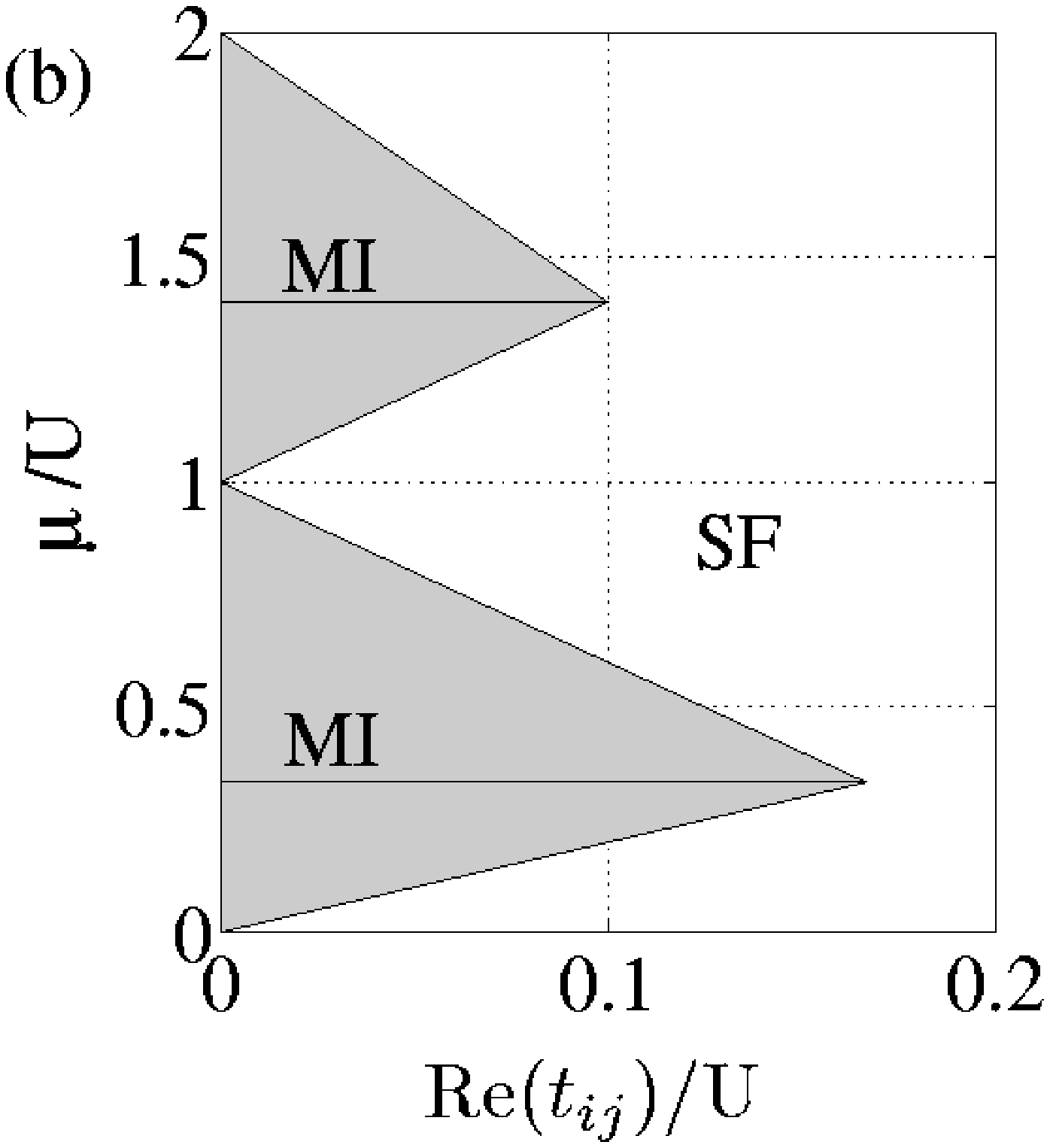} 
          \includegraphics[width=4cm]{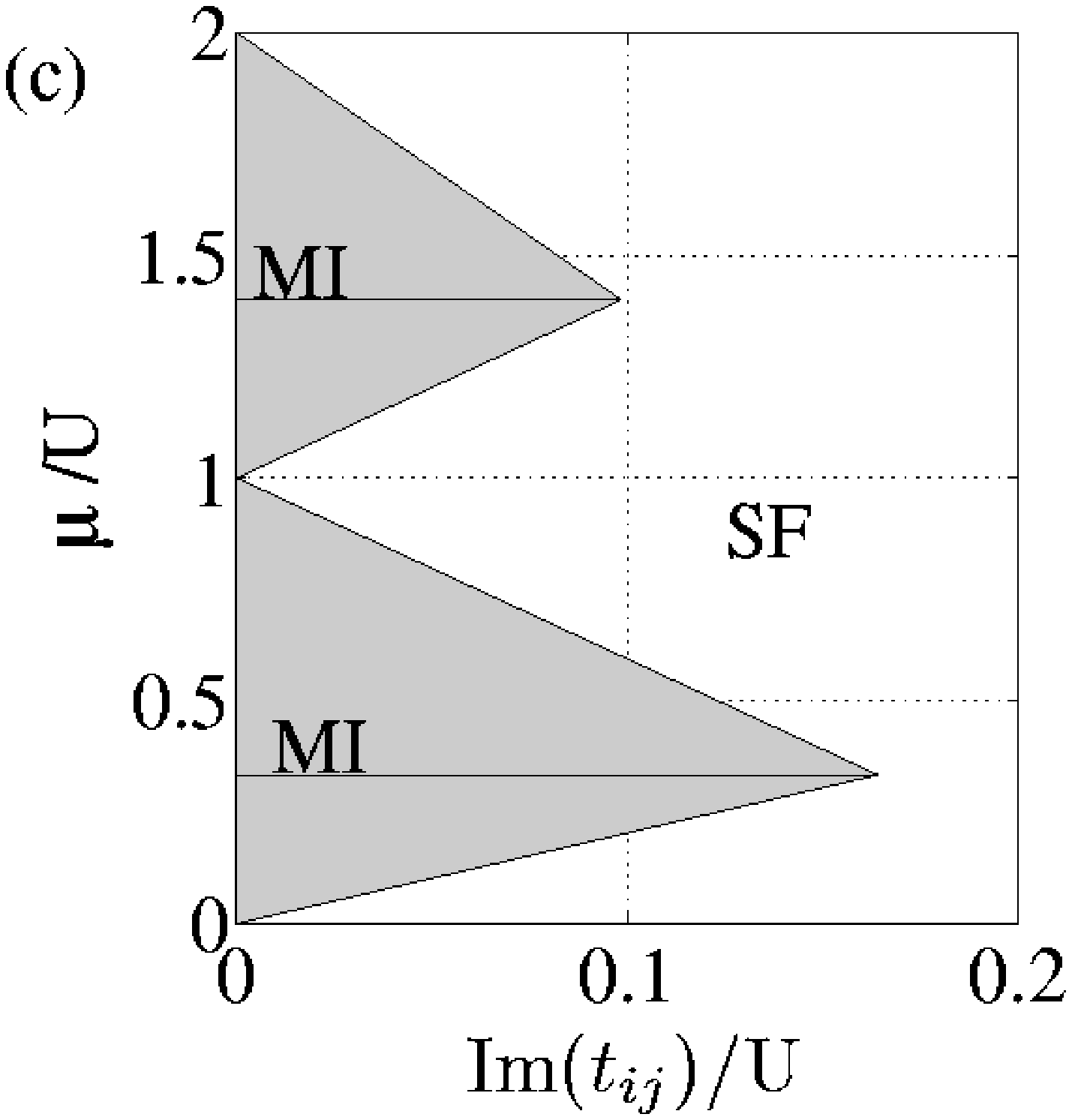}   
  \end{minipage}
  \caption{ (a) Mott-Insulator zone boundaries for bosons in a
    $2\times 2$ lattice as a function of chemical potential $\mu$ and
    of the real and imaginary parts of the complex hopping parameter
    described in Eq.~(\ref{tij}). The volume inside the shaded
    surfaces corresponds to Mott-insulating states with average
    fillings of one and two particles per site in the lower and upper
    volume respectively.  (b) The $x-z$ cross-section of the phase
    diagram is the same as that obtained using the Bose-Hubbard
    Hamiltonian for a non-rotating system. (c) The $y-z$ cross section
    of the phase diagram is similar to the $x-z$ cross section.  Note
    that the shape of the Mott zones will change on going beyond the
    two-state approximation~\protect{\cite{note:twostate}}.
    \label{PhaseDiagram}}
\end{center}
\end{figure}

The phase diagram for the non-rotating Bose-Hubbard model is obtained
in the grand-canonical ensemble by adding a term $-\mu\sum_i {\hat
  n}_i$ to the Hamiltonian. The phase diagram separates into two
regions---the Mott insulator (MI) lobes with commensurate filling
(integer number of atoms per site) and the superfluid (SF) regions
with incommensurate filling. The phase diagram for the non-rotating
lattice was first studied in Ref.~\cite{Fisher:1989}.

The introduction of rotation makes the hopping energy parameter $t$
complex and site-dependent as can be seen by rewriting Eq.~(\ref{MBH})
in the grand canonical description after combining the first and last
terms.
\begin{eqnarray}
\hat{H}&=&- \! \sum _{\langle i,j \rangle} \left( t _{ij}\hat{a} _i
^{\dagger} \hat{a} _j +t^* _{ij}\hat{a}_i \hat{a} _j ^{\dagger}\right)
+(\epsilon-\mu) \sum _i \hat{n} _i \nonumber \\ &&+ \frac{U}{2}\sum _i
\hat{n}_i (\hat{n}_i - 1), \label{tic}
\\ t_{ij}&=& t+i\hbar \Omega K
_{ij}. \label{tij}
\end{eqnarray} 
The surface in Fig.~\ref{PhaseDiagram} represents the boundary between
the Mott insulator (MI) and superfluid (SF) regions as a function of
the chemical potential and of the real and imaginary parts of the
hopping parameter $t _{ij}$. The zero point energy $\varepsilon$ in each 
well has been set to zero since it gives rise to an
irrelevant overall shift. Note that the shape of the tip of the Mott zones 
in Fig.~\ref{PhaseDiagram} differs slightly if more than two Fock states are 
used to represent each state~\cite{note:twostate}.  
At $\Omega=0$ ($\mathrm{Im}(t_{ij})=0$, Fig.~\ref{PhaseDiagram}(b)), 
the phase boundary matches that for the
non-rotating Bose-Hubbard model in the two-state approximation. A
similar diagram is obtained when setting $\mathrm{Re}(t_{ij})=0$
(Fig.~\ref{PhaseDiagram}(c)).  In general, this similarity need not be
present for larger systems where $t _{ij}$ is not the same for all
pairs of nearest neighbors as in the example of a $2\times 2$ lattice.
Note that even though $\mathrm{Im}(t_{ij})$ can be varied freely via $\Omega$
for a given $t$ and non-zero $K_{ij}$, the reverse is not true. This
is because for a particular realization of a lattice, the two overlap
integrals $t$ (Eq.~(\ref{t})) and $\beta$ (Eq.~(\ref{beta})) are both
fixed by the lattice depth. Hence, a fixed $\mathrm{Im}(t_{ij})=\Omega K_{ij}$
implies a fixed $\mathrm{Re}(t_{ij})=t$. In the case of a standing wave optical
lattice in the tight binding regime, $K_{ij}$ is approximately
proportional to $t/E_R$ (see Fig.~\ref{Overlap}). This makes the plane
$\mathrm{Re}(t_{ij})=0$ of the phase diagram inaccessible.

\subsection{Differences between even and odd lattices}

\begin{figure}[t]
\begin{center}
   \includegraphics[width=7.8cm]{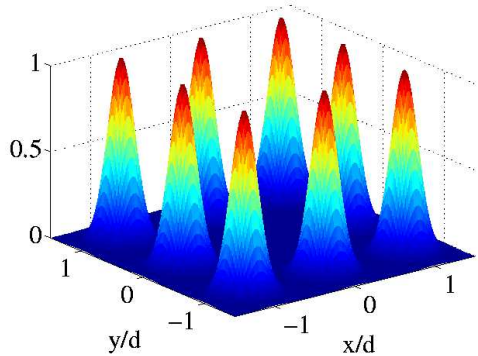}
   \caption{ Number density distribution for one particle in a
   $3\times 3$ lattice of depth $V_0=10E_R$ at a lattice rotation of
   $\hbar\Omega=E_R$. There is almost complete number depletion in the
   central site because it coincides with the center of rotation and
   hence the vortex core.
\label{CenterDepletion}}
\end{center}
\end{figure}

In the study of rotating lattices, the position of the center of
rotation gives rise to an important distinction between two kinds of
lattices: lattices with an even number of sites (e.g., $2 \times
2$, $4 \times 4$) and lattices with an odd number of sites (e.g., $3 \times 3$, 
$5 \times 5$).  If assumed to be at the center of the
system, the axis of rotation passes through a peak in the lattice
potential in the case of even lattices while it passes through the
central site for odd lattices.  It is useful to briefly touch upon two
interesting aspects of the ground state solution for odd lattices: (1)
there is nearly complete number density depletion in the center site
when rotation enters the system, as would be expected at the vortex
core in a continuous system (Fig.~\ref{CenterDepletion}); and (2)
there are no currents along the radial direction ($K_{ij}=0$) so that
particles only hop along the azimuthal direction.  Even lattices have
no sites which are nearest neighbors in a strictly radial direction,
but the main results for particles in a rotating even lattice can be
mapped onto those for particles in an odd lattice. Only the even
lattice is discussed for the rest of the paper since it captures the
most important physical features.

\subsection{Number density distribution}

\begin{figure}
  \centering
  \begin{minipage}{4 cm}
    \includegraphics[width=4cm]{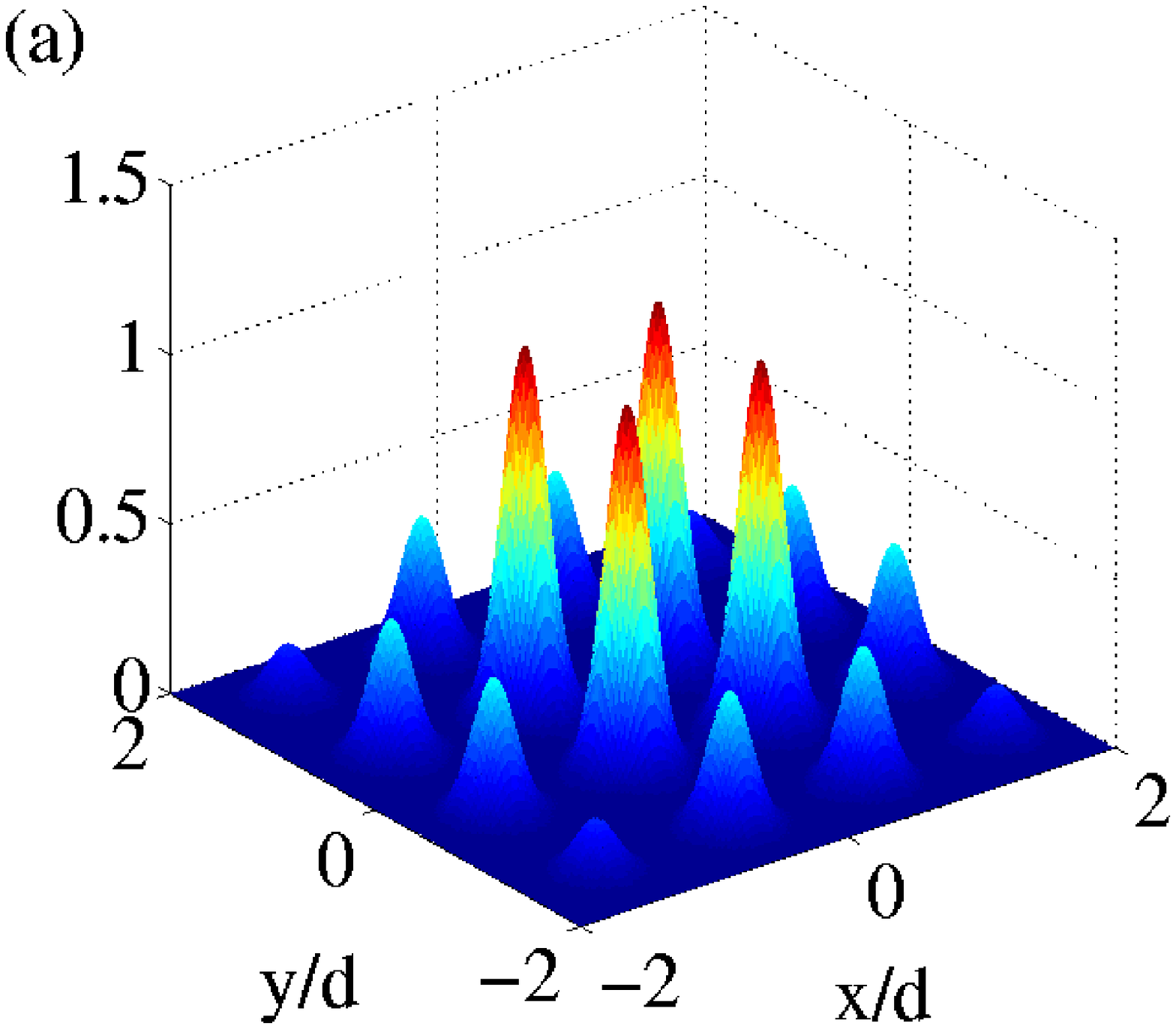} 
     \includegraphics[width=4cm]{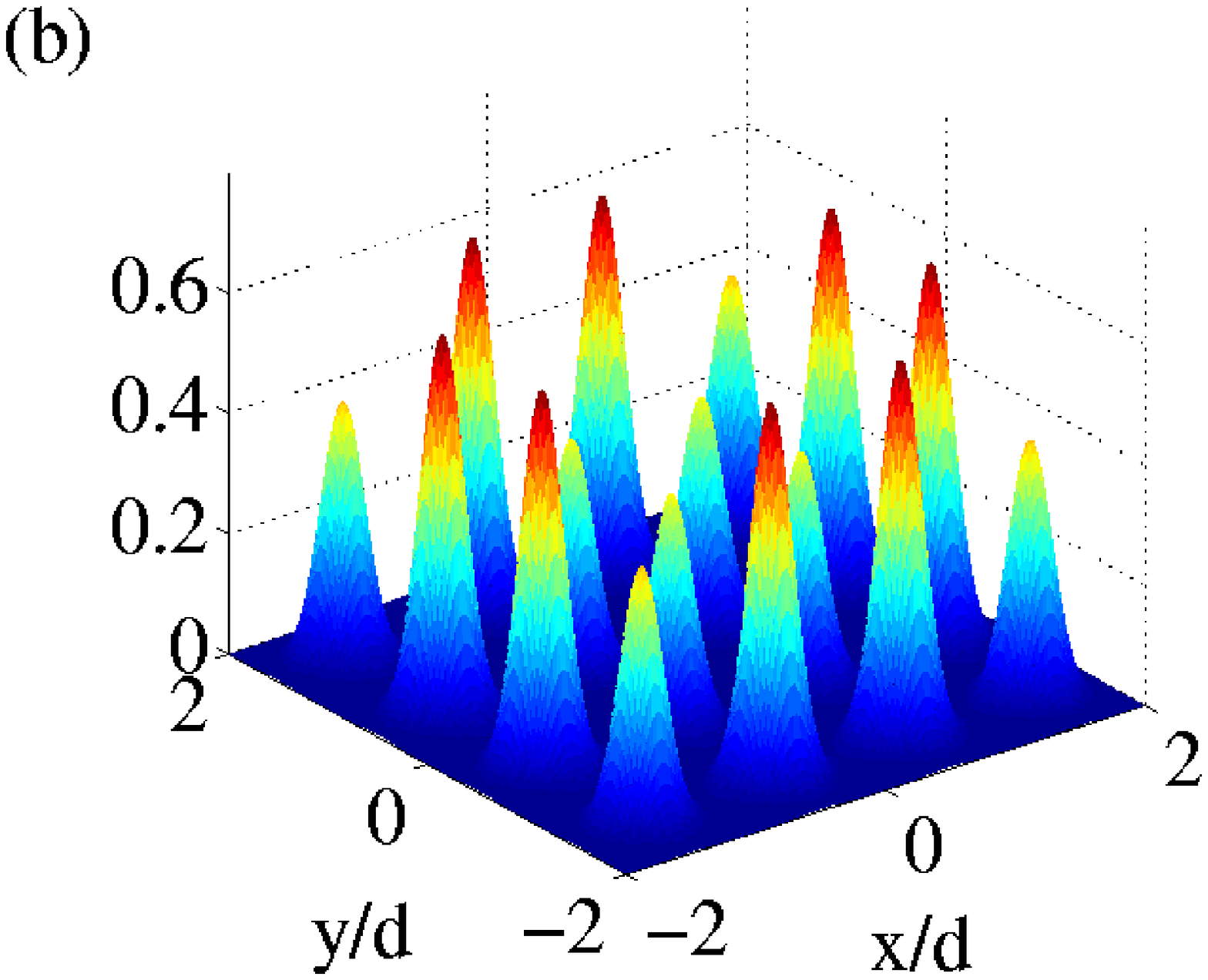}   
  \end{minipage}
  \begin{minipage}{4.5 cm}
    \includegraphics[width=4cm]{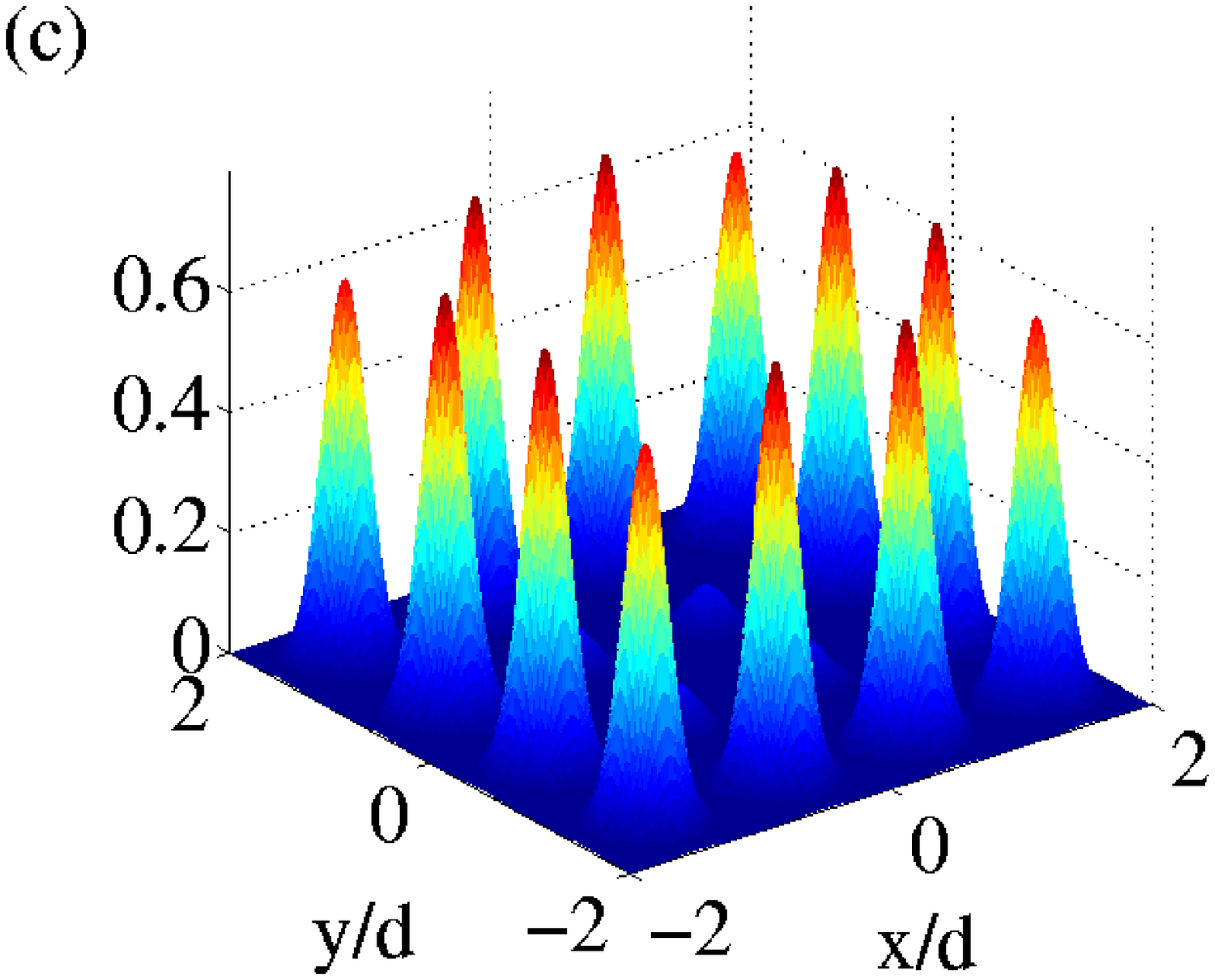} 
     \includegraphics[width=4cm]{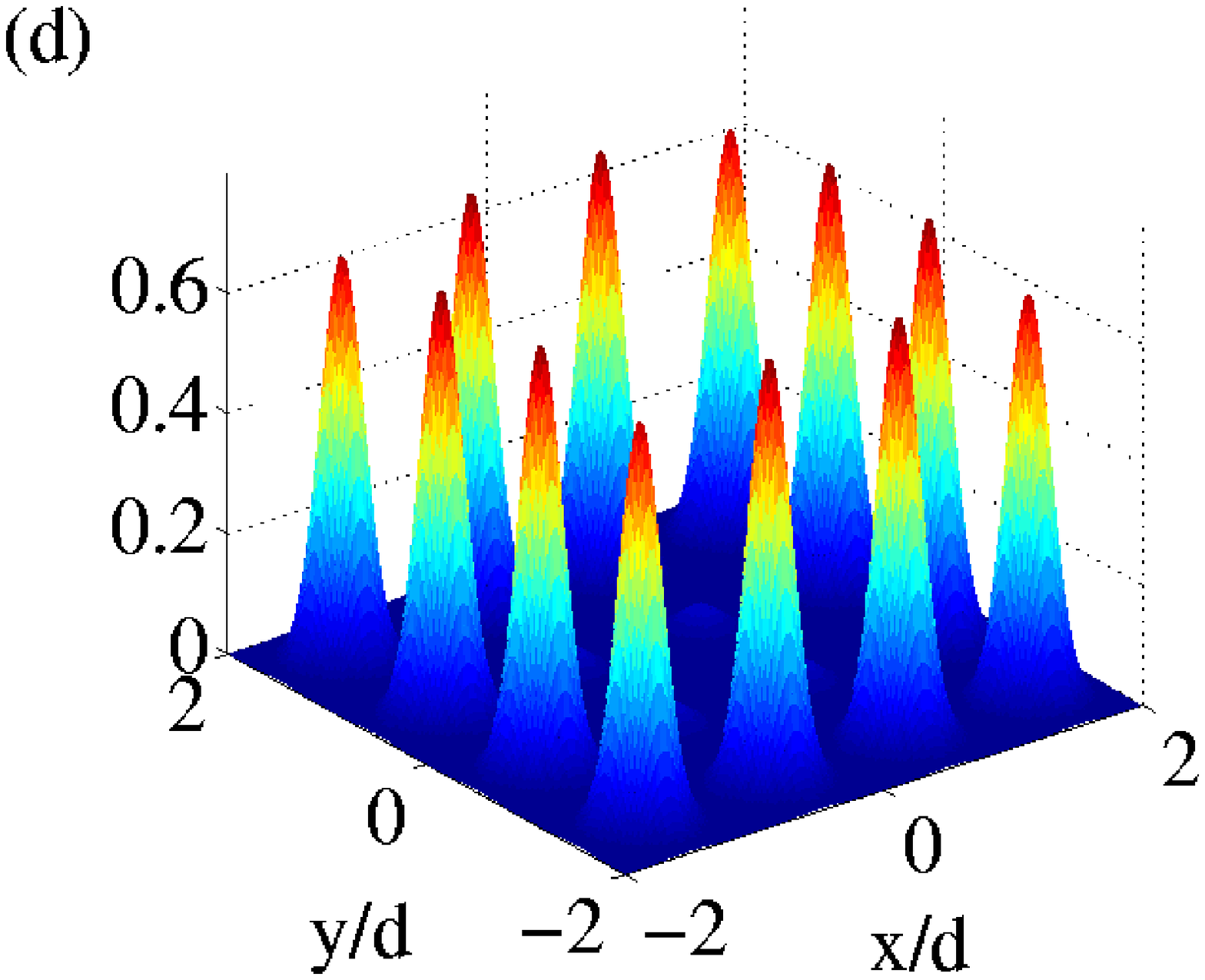} 
  \end{minipage}
  \caption{Number density distribution for one particle in a
    $4\times4$ lattice at $V_0=10 E _R$. (a) $\hbar\Omega=0.1 E_R$ -
    rotation has yet to enter the system. The center-peaked
    distribution is due to the infinite potential walls at the
    perimeter of the lattice. (b) $\hbar\Omega=0.2 E_R$ (c)
    $\hbar\Omega=0.4 E_R$ (d) $\hbar\Omega=0.8 E_R$. At large
    rotation, particles get pushed to the outermost sites creating a 
    washing machine effect.}
  \label{NumberDensity}
\end{figure} 

The number density distribution is obtained by evaluating the
expectation of the site-specific number operator $\hat{n}_i$.
Figure~\ref{NumberDensity} describes the number distribution for one
particle in a $4 \times 4$ lattice with phase windings of 
0, $2\pi$, $4 \pi$, and $6\pi$.  For small angular velocities, there
is no effect of rotation on the ground state of the system and the
number density is center-peaked (Fig.~\ref{NumberDensity}(a)). In the
continuous limit of the lattice spacing becoming infinitesimal, this
mirrors the number distribution for particles in a 2D box with
infinite potential walls. The distribution changes each time a vortex
enters the system.  For large rotation (Fig.~\ref{NumberDensity}d), 
we observe a {\it washing machine   effect}, as 
the number density gets concentrated at the perimeter. The inner sites 
have a small non-zero number density in this limit.


\section {Quasi-angular Momentum \label{Section:QAM}}

The presence of the lattice breaks the continuous rotational symmetry
of the system. The eigenvalues of the angular momentum operator are
therefore no longer good quantum numbers because the rotational symmetry
associated with $\hat{L}$ has been replaced with a discrete rotational
symmetry. In this section, ideas of discrete translational symmetry
and Bloch's theorem are mapped onto a discrete rotational symmetry
problem to generate quasi-angular momentum states. Exact results are
presented for the modified Bose-Hubbard model in the context of a
single particle one-dimensional ring, and connections are made with
the square lattice.

\begin{figure}[t]
\begin{center}
   \includegraphics[width=8.6cm]{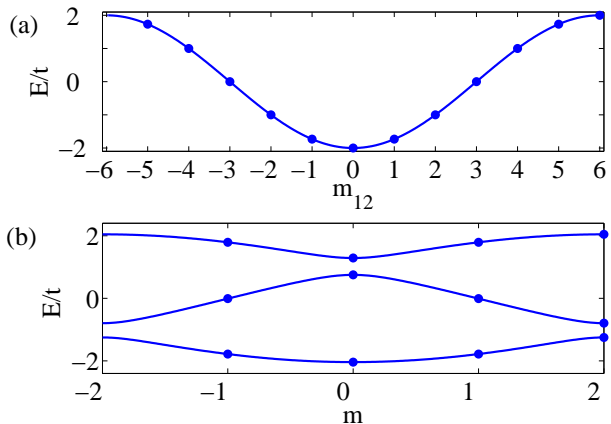}
   \caption{(a) Energy-quasi angular momentum dispersion
   relationship for a 12-site ring. Positive and negative $m _{12}$
   values indicate rotation in opposite directions. The velocity of
   the particles in any state is given by the slope of the dispersion
   curve at the point. Accordingly, the velocity is maximum at $m=3$
   (b) Energy---quasi-angular momentum dispersion relationship for a
   12-site ring with a small four-fold periodic potential. Note the lines joining
   states are obtained by extrapolating for an infinite system.
   \label{BlochBand}}
\end{center}
\end{figure}

Consider a one-dimensional lattice with periodic boundary
conditions, i.e., a ring lattice of $N$ sites. A rotation of
$2\pi /N$ leaves the system invariant and hence the rotation operator
${\rm R}\left( 2\pi/N\right) $ commutes with the Hamiltonian. This is
also true for a square ring because the site dependent parameter $K
_{ij}$ in the Hamiltonian (Eq.~(\ref{MBH})) depends on the
perpendicular distance of the line connecting two nearest neighbor
sites from the center of rotation. The energy eigenstates can be
labeled using the eigenvalues of ${\rm R}\left( 2\pi/N\right)$:
\begin{equation}
R\left( 2\pi /N\right) \left\vert m \right\rangle =e^{i2\pi
m/N}\left\vert m \right\rangle :m\in \{0,\dots ,N-1\}, \label{DSR}
\end{equation}
where the exact eigenvectors shown can easily be derived by expanding $
\left\vert m \right\rangle $ in the Fock basis and demanding periodic
boundary conditions. 

At this point, it is useful to make a connection with
conventional Bloch theory. ${\bf R}(\pi/2)$ is analogous to the
discrete translation operator ${\bf T}(d)$ for a stationary
one-dimensional lattice of period $d$~\cite{Ashcroft},
\begin{eqnarray}
{\bf T}(d)\Psi (x_1,\ldots,x_n) = e^{iqd}\Psi (x_1,\ldots,x_n), \label{T}
\end{eqnarray}
where $\Psi(x_1,\ldots,x_n)$ is an eigenfunction of the translation operator. 
The eigenvalues of ${\bf T}(d)$ are described by the quasi-momentum $q$. 
In a way exactly analagous to that of quasi-momentum Bloch states for a 
discrete translation operator, we can identify these $m$-values in Eq.~(\ref{DSR}) 
as quasi-angular momenta.  Note that this discussion so far is completely
general and applies to both the single-particle and many-particle
cases.

To illustrate the role of the quasi-angular momentum and the
connection with the quasi-momentum in systems with discrete
translational symmetry, consider one particle in a $12$ site static
ring. Each of the sites is indexed by an azimuthal coordinate, $\phi
_i$. The energy spectrum takes on the well-known dispersion relation
observed for the lowest Bloch band of a particle in a 1D lattice with
periodic boundary conditions in the tight-binding regime
(Fig.~\ref{BlochBand}(a)). Since the system has 12-fold symmetry, the
quasi-angular momentum, $m$, can take on 12 possible values. The slope
of the energy plot provides the velocity of the particle. As discussed
before (Section \ref{Section:Hamiltonian}), rotation is introduced by
adding a term, $- (\hbar \Omega/i)\partial _\phi$, to the Hamiltonian
in order to obtain the ground state in rotating frame coordinates. As
$\Omega$ is ramped in a particular direction, the ground state
quasi-angular momentum changes from $m=0$ to $m=3$ in steps of 1 (not
shown here). The $m=3$ state corresponds to the maximum slope of the
dispersion and the largest particle velocity.  A particle in the $m=4$
state has the same velocity as the $m=2$ state but with a higher
energy. Quasi-angular momenta  $m=7,\ldots,11$ correspond to $m=-5,\ldots, -1$ and describe circulation in the opposite direction. This is described by the $C_{12}$ point symmetry group. 

Consider now a 12-site ring perturbed by a four-fold symmetric periodic potential. The Hamiltonian, $H=H_{12}+V$, is the sum of two terms, the 12-site 
Hamiltonian $H_{12}$ which has a 12-fold rotation symmetry and a 
potential $V$ which has a four-fold rotation symmetry. Figure~\ref{BlochBand}(b) 
is the energy dispersion relation as a function of quasi-angular momentum for
the 12-site lattice ring with this small four-fold symmetric potential. Since the potential
increases the rotational symmetry from $d=2\pi/12$ to $d=2\pi/4$, the Brillouin zone
is narrowed down to $m=-2,\ldots,2$. Three energy bands are created in place of one. 
States on adjacent bands with the same $m$ value --- for example $m_{12}=-2$ and $m_{12}=2$ --- are mixed by the four-fold symmetric potential $V$, thereby leading to an energy gap at this $m$ value.

An analogous situation occurs when we try to qualitatively understand the properties 
of a $4\times 4$ lattice, which has 12 sites on the boundary. We adopt a perturbative approach by breaking the system into two non-interacting 12-site and 4-site rings and considering an interaction between them, i.e., $H=H_{12}+H_{4}+V$. The interaction 
with the four-site ring breaks the 12-fold symmetry of the outer ring, reducing it to a 
four-fold discrete rotational symmetry. 

The above example illustrates that a particle in a square lattice is
characterized by a four-fold discrete rotational symmetry.  The same
symmetry considerations hold for many particles in the system. Hence,
the many-body eigenstates are quasi-angular momentum states with
$m\in \{-2,1,0,1,2\}$.

In the following we show how the rotation of the lattice leads to a
change in quasi-angular momentum in the groundstate of the system in
the single-particle case (Section~\ref{Section:SingleParticle}) and in
the many-body case (Section~\ref{Section:ManyParticles}). In addition, we
show how these transitions affect other properties of the system, such
as its average angular momentum and its vorticity. 

\section{One-particle analysis \label{Section:SingleParticle}}

This section examines the response of one particle to lattice rotation. 
The advantage in first considering only one particle is that it allows 
one to distinguish general characteristics of
the systems from effects due to interaction.

\begin{figure}[t]
  \centering
    \includegraphics[width=8.4cm]{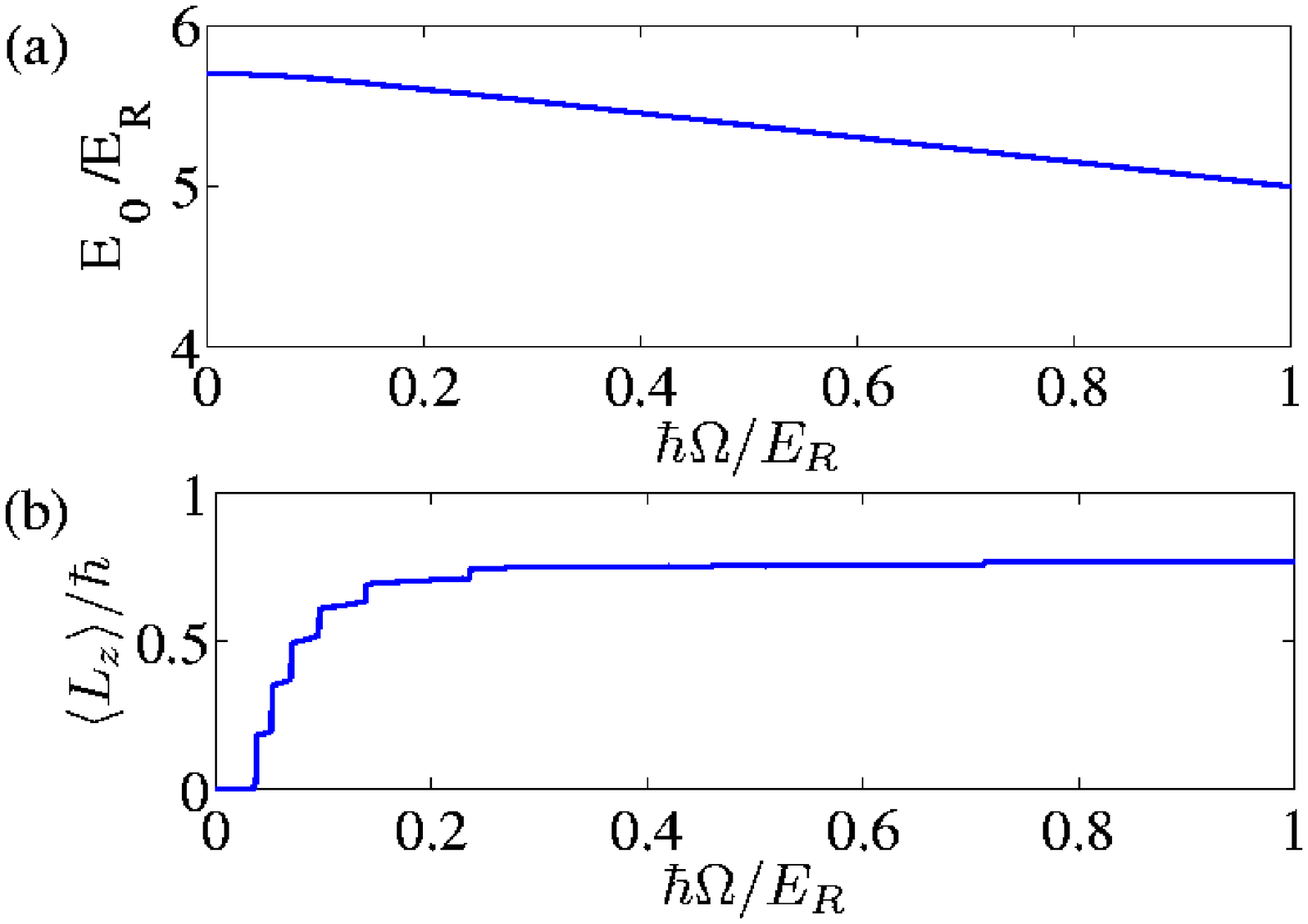}  
    \includegraphics[width=8.4cm]{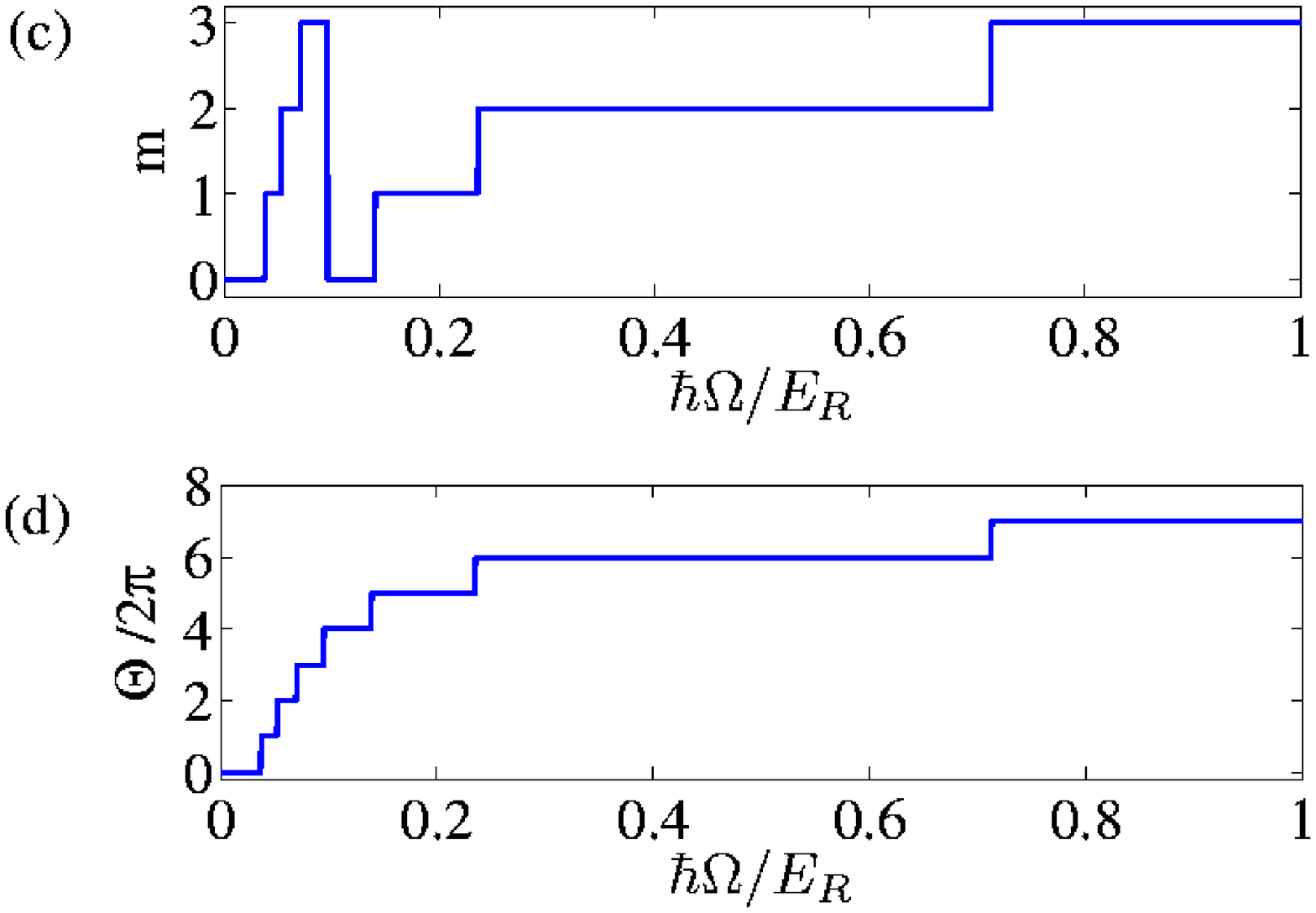}
  \caption{One particle in an $8 \times 8$ lattice.  (a) Ground state
  energy, $E_0$ vs. $\Omega$. (b) Average angular momentum,
  $\langle L_z\rangle$ vs $\Omega$. Note that the expectation value of
  angular momentum (see Eq.~(\ref{L})) is not quantized. (c) Quasi-angular 
  momentum $m$ vs. $\Omega$. In direct analogy to quasi-momentum values 
  for linear lattices, $m$ repeats itself. (d) The phase winding, 
  $ \Theta /2\pi$, vs. $\Omega$.  For the $8\times8$ lattice, the maximum phase winding is $7\times 2\pi$. 
  \label{SingleParticle}}
\end{figure} 

Figure~\ref{SingleParticle} describes the response of the system as a
function of the angular velocity $\Omega$.
Figures~\ref{SingleParticle}(a) and \ref{SingleParticle}(b) show that
for increasing $\Omega$ the ground state energy in the rotating frame
$E _0$ decreases with discontinuous derivative as different states
become energetically favorable. Note that the fact that the
eigenstates are not eigenstates of angular momentum is explicitly
illustrated here since $\langle L_z \rangle$ takes on non-quantized
values.

The abrupt changes in average angular momentum are connected to
changes in the quasi-angular momentum $m$ of the groundstate as seen
in Fig.~\ref{SingleParticle}(c). Since the lattice has four-fold
rotational symmetry, the values that the quasi-angular momenta can
take on are $m\in\{0,1,2,3\}$. For additional transitions, $m$ repeats
itself in behavior analogous to that of {\it linear quasi-momentum} as
one crosses the first Brillioun zone. 

The changes in quasi-angular momentum are associated with changes in
the phase winding of the single particle wavefunction.  The phase
winding $\Theta$ jumps by $2 \pi$ each time the quasi-angular
momentum of the groundstate changes (Fig.~\ref{SingleParticle}(d)).
The maximum phase winding of $14\pi$ for an $8\times8$ lattice
corresponds to a maximum phase difference of $\pi/2$ between any two
sites on the lattice boundary. A difference of $\pi/2$ between two
lattice sites corresponds to the condition for the maximum current
attainable within the lowest band Bose-Hubbard model.  This result can
be generalized to a lattice of size $L\times L$. The number of sites
on the circumference of the lattice is $4(L-1)$ and for a
phase difference of $\pi/2$ between two adjacent perimeter sites, the maximum 
phase winding around the circumference is $2\pi(L-1)$ within the lowest
band. Since the notions of an order parameter and of superfluidity do
not apply to single particle systems, states with non-zero phase
winding can not be referred to as quantized vortices. However, as will
be shown in the following, these single particle results extend to 
many-particle systems in a straightforward manner.

\section{Many-particle analysis\label{Section:ManyParticles}}

This section probes the effects of strongly repulsive interaction
 when the number of particles in the system is greater than
one. The symmetry considerations discussed in
Section~\ref{Section:QAM} are first tested for many particles. The
effect of a symmetry-commensurate filling is explored by considering
two different systems - (1) four particles in a $4\times4$ lattice and
(2) five particles in a $4 \times 4$ lattice. These two systems
correspond to fillings commensurate and incommensurate with the
four-fold symmetry of the lattice, respectively.

\begin{figure}[t]
\begin{center}
   \includegraphics[width=8.6cm]{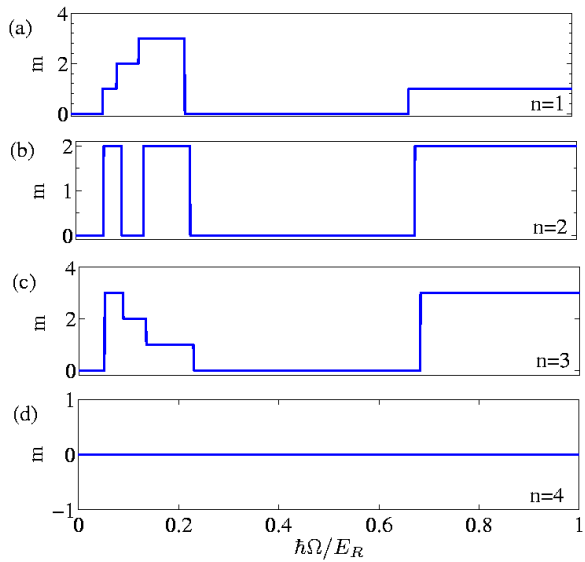}
   \caption{Quasi-angular momenta for one through four strongly
   repulsive bosons in a $6\times 6$ lattice. For multiple particles in 
   the lattice and increasing
   rotation, the quasi-angular momenta cycles through values given by
   $m=nl ~{\rm mod}~ 4, l=0,1,2,3$, where $n$ is the number of
   particles. (a) $n=1:m=0,1,2,3,0,1$ (b) $n=2: m=0,2,0,2,0,2$ (c)
   $n=3: m=0,3,2,1,0,3$ (d) $n=4: m=0$. \label{QAM} }
\end{center}
\end{figure}

As discussed in Section~\ref{Section:QAM}, the four-fold rotational
symmetry allows labeling of the many-body states by their
quasi-angular momenta. In a static lattice, the groundstate is always
characterized by $m=0$. This value may change when the lattice is
rotated at angular velocity $\Omega$, as has been demonstrated in the
single-particle case. In contrast to the single-particle case, the
many-body groundstate does not necessarily cycle through all possible
$m$-values as $\Omega$ is increased. Instead, the values of quasi-angular
momenta it can take on depend on the number of particles.  As illustrated in 
Fig.~\ref{QAM}, we see that for $n$ particles in the system, the
quasi-angular momentum of the ground state cycles through values
satisfying the relation
\begin{equation}
m=nl~{\rm mod}~4\,,
\label{QAMcycle}
\end{equation}
where $l\in \{0,1,2,3\}$. The validity of this expression has been verified
both numerically for various lattice sizes and particle numbers and
analytically within a Jordan-Wigner transformation approach to
hard-core bosons in a ring~\cite{Peden:TBP}.  Hence, only for
odd $n$ does the quasi-angular momentum of the groundstate
cycle through all values of $m$. Cases in which the particle number is
commensurate with the four-fold symmetry $n=4,8,\ldots$ are of
particular interest in that these systems always stay in an
$m=0$ state. Note that a simplistic explanation for the validity of 
Eq.~(\ref{QAMcycle}) is obtained if all particles occupy a condensate 
mode with quasi-angular momentum $m=0,1,\ldots$, yielding a total
quasi-angular momentum equal to 0 when $n=4$.

\subsection{Symmetry-commensurate filling}

\begin{figure}[t]
  \centering
    \includegraphics[width=7.8cm]{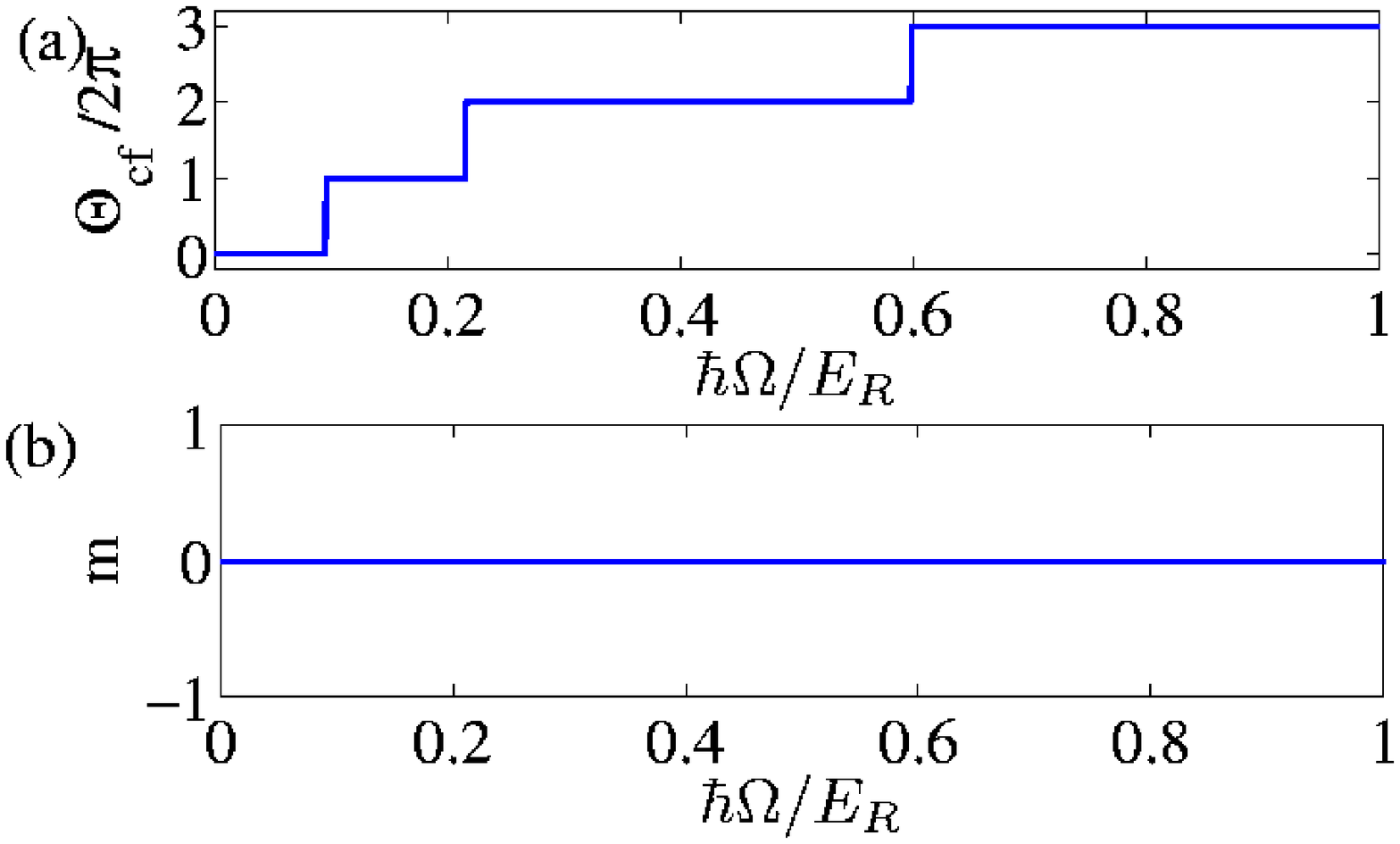}  
     \includegraphics[width=7.8cm]{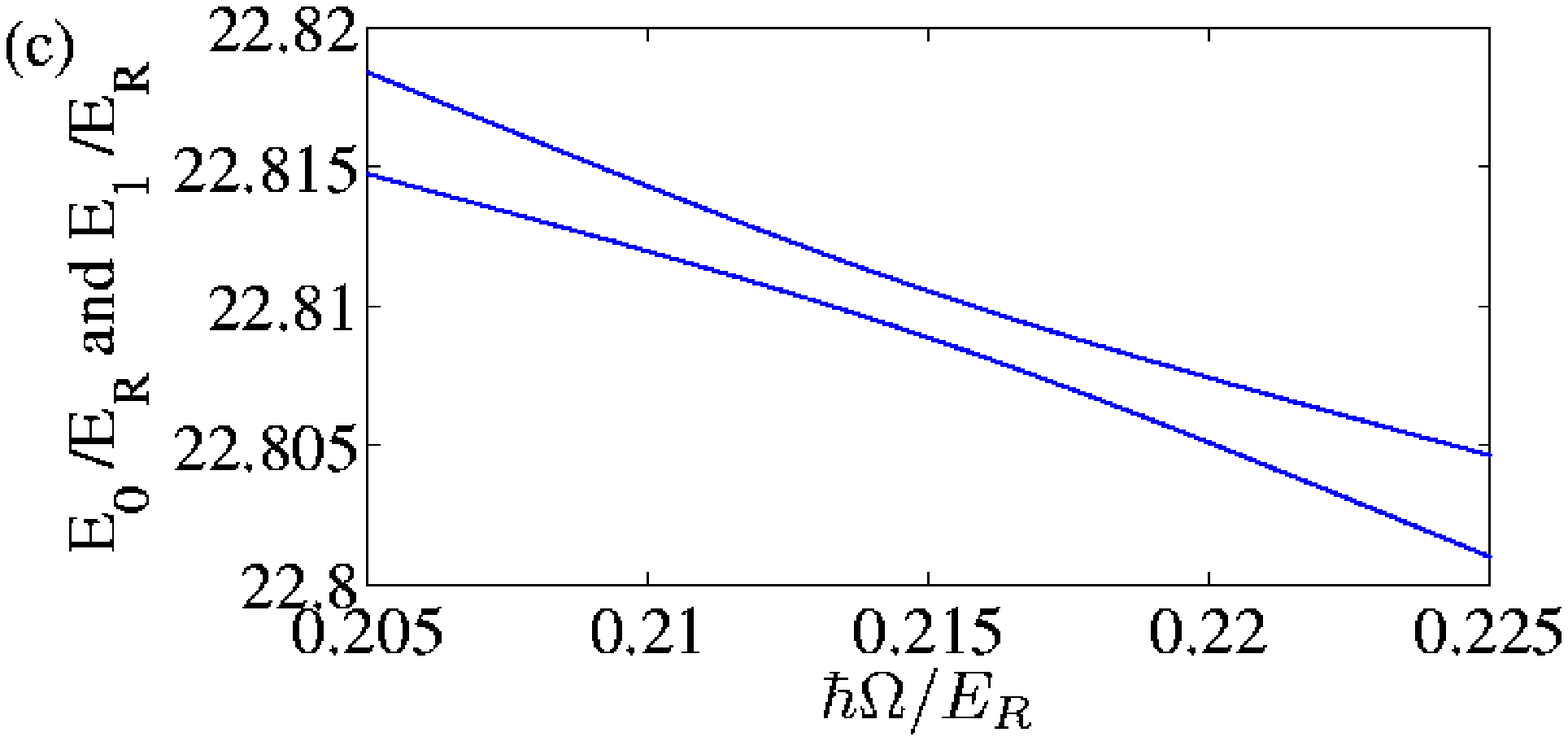}
  \caption{Four particles in a $4 \times 4$ lattice. (a) Number of
  vortices, $\Theta _{cf}/2\pi$, vs. $\Omega$. Three vortices enter the
  $4\times 4$ lattice. (b) Quasi-angular momenta, $m$ vs. $\Omega$. The
  symmetry of the ground state as indicated by the quasi-momentum $m=0$
  does not change even with three vortices entering the system. (c)
  Zoom-in of lowest two energy levels around the entry of the second
  vortex shows an avoided energy level crossing. The mixing of states is possible
  because the ground states on either side have the same discrete 
  rotational symmetry.}
 \label{AvoidedCrossing}
\end{figure} 

When the number of particles is commensurate with the four-fold
rotational symmetry, the groundstate always has zero quasi-angular
momentum. This does not exclude the entry of quantized vortices into
the system. To give an example, we analyze the case of four particles
in a $4\times 4$ lattice. The largest eigenvalue of the ground state one-body
density matrix is found to be $60-74\%$ of the total particle number. 
Since all other eigenvalues are significantly smaller, this number is large 
enough to refer to the corresponding
eigenmode as the condensate wavefunction. The phase winding
$\Theta_{cf}$ of the condensate wavefunction increases in steps of
$2\pi$ to a maximum of $6\pi$ as the lattice is rotated faster and
faster (see Fig.~\ref{AvoidedCrossing}(a)). This corresponds to a
maximum of $L-1=3$ quantized vortices, with $L\times L$ being the size
of the lattice. As in the single particle case, the maximum phase winding 
that can be observed within a lowest band model is limited by the
maximum phase difference of $\pi/2$ between neighbouring sites.

Each vortex entry is associated with an avoided crossing between
groundstate and first excited state. This is possible because both states 
have quasi-angular momentum, $m=0$, allowing them to mix around the 
vortex entry point. This is demonstrated in
Fig.~\ref{AvoidedCrossing}(b) for the entry of the second vortex.

\subsection{Symmetry-incommensurate filling}

\begin{figure}[t]
  \centering
  \begin{minipage}[c]{7.8cm}
    \includegraphics[width=7.8cm]{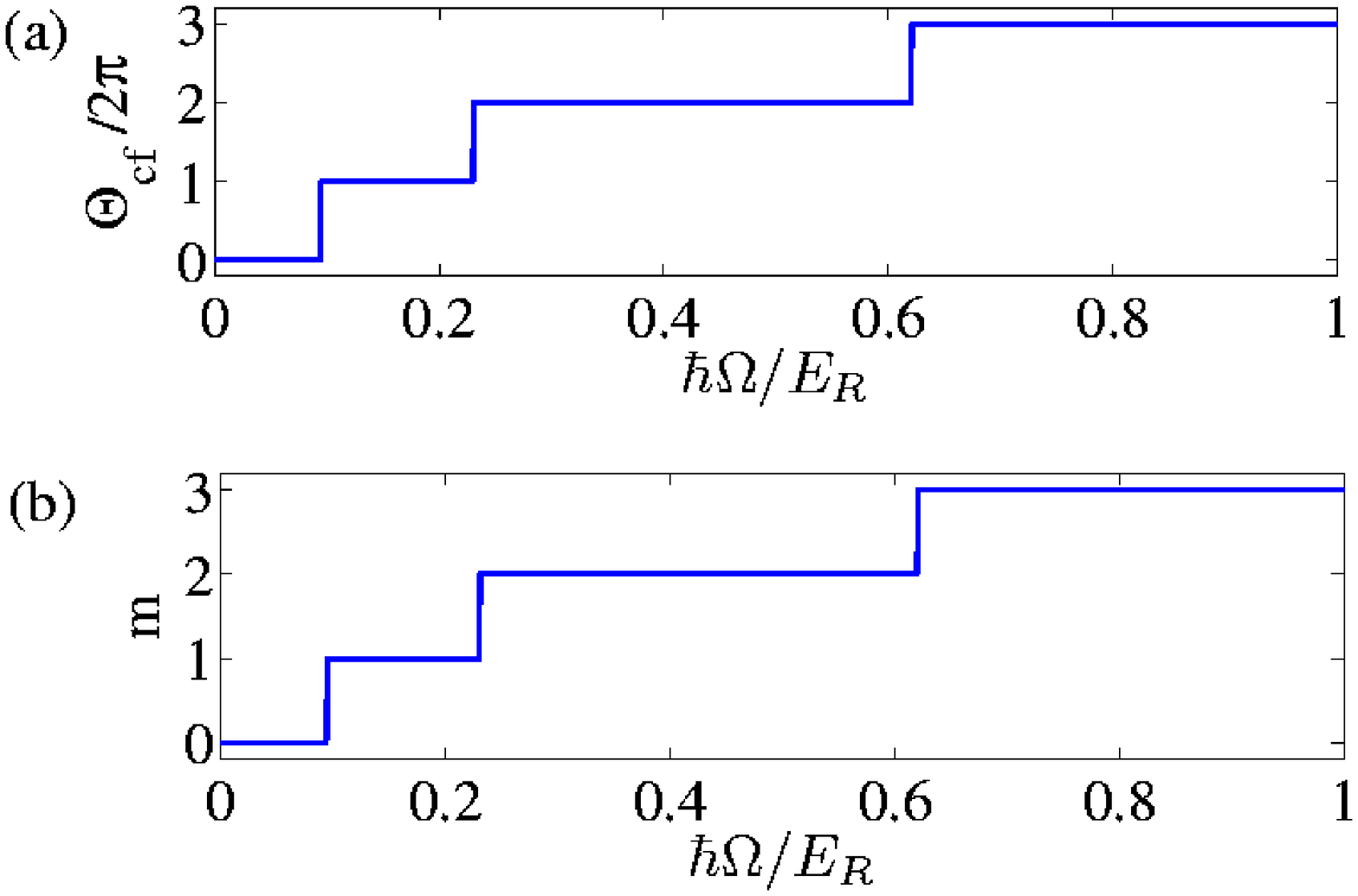} 
    \includegraphics[width=7.8cm]{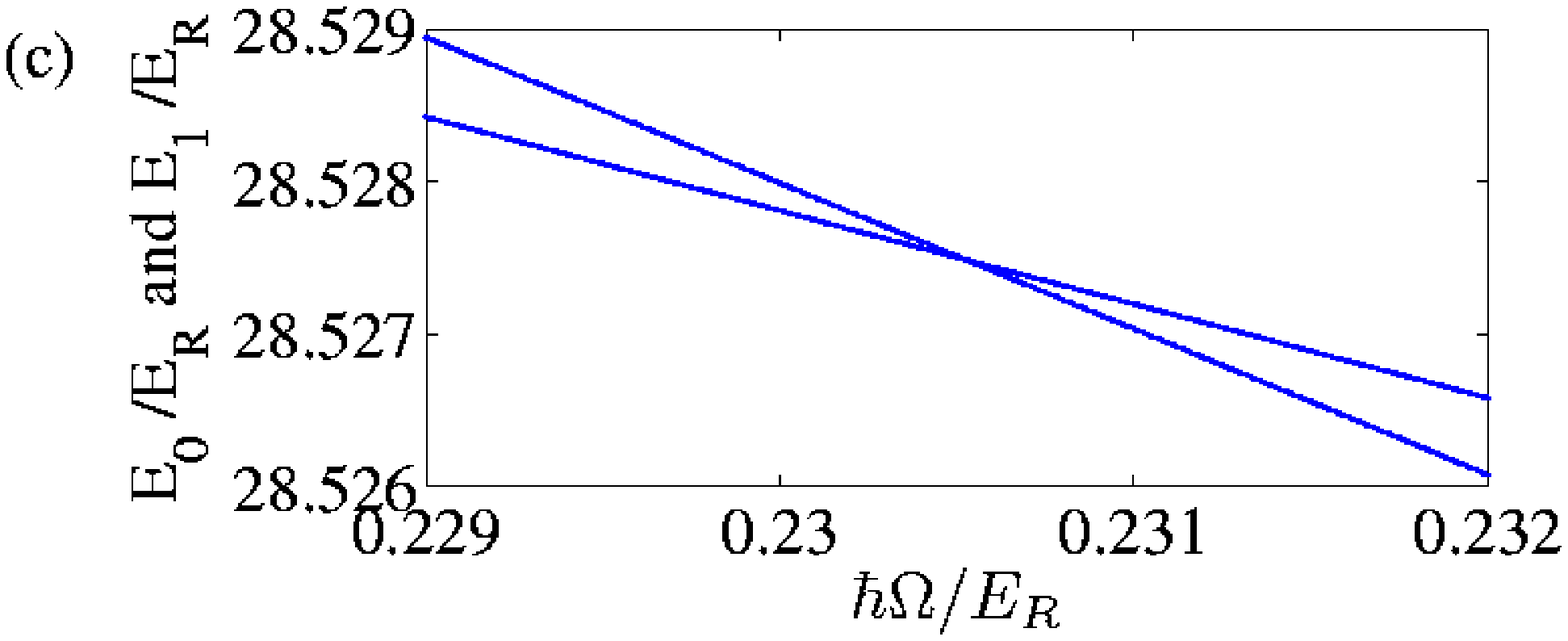} 
  \end{minipage}
  \caption{Five particles in a $4 \times 4$ lattice. (a) Number of
  vortices $\Theta _{cf}/2\pi$ vs. $\Omega$. As before, a maximum of
  three vortices enter the system (b) Quasi-angular momenta $m$
  vs. $\Omega$. Since the filling is incommensurate with the
  symmetry, the quasi-angular momenta takes on values $m=0,1,2,3$ (Eq.~\ref{QAMcycle}). (c) Zoom-in of lowest two energy levels around the entry of the second
  vortex shows an energy level crossing as the ground state symmetry
  changes as a function of a parameter in the Hamiltonian, $\Omega$.}
  \label{Crossing}
\end{figure} 

In behavior similar to that for four particles, the eigenmode corresponding to the condensate wavefunction is macroscopically occupied, and $\Theta _{cf}$ for a
symmetry-incommensurate number of particles increases in steps of
$2\pi$ up to a maximum of $2\pi(L-1)$. Yet, in contrast with a
symmetry-commensurate filling, the discrete rotational symmetry of the
system changes with each vortex entry.  This is demonstrated in
Fig.~\ref{Crossing}(a) and (b) where we plot both the phase winding
and the quasi-angular momentum as a function of $\Omega$ for the case
of five particles in a $4\times 4$ lattice.  In this setting, the
maximum phase winding is given by $6\pi$ while the quasi-angular
momentum takes the values $m=0,1,2,3$ in accordance with
Eq.~(\ref{QAMcycle}). 

Since for symmetry-incommensurate filling the symmetry of the
many-body wavefunction is different on either side of the jump in the
phase winding, transitions between vortex states cannot occur via the
mixing of energy eigenfunctions with the same symmetry. Hence, changes
in vorticity are not associated with an avoided crossing between the
ground state and the first excited state. Instead, the transition occurs as
the energy of an excited state with different quasi-angular momentum
and phase winding drops low enough to become the new groundstate. The
signature of vortex entry is thus a crossing of energy levels with
different discrete rotational symmetry and phase
winding. Fig.~\ref{Crossing}(c) depicts the level crossing associated
with the entrance of a second vortex into a system of five particles
in a $4\times 4$ lattice.  The level crossings are a non-trivial
result for many particles since they correspond to a symmetry change 
in the ground state as a function of a parameter of the Hamiltonian and 
are indicative  of quantum phase transitions~\cite{Sachdev:2001}.

\section{Conclusions\label{Section:Conclusions}}

We have studied zero temperature hard-core bosons in 2D rotating
square lattices for filling factors of less than one atom per site
using a modified Bose-Hubbard Hamiltonian. An important feature of the
system is the quasi-angular momentum, reflecting the discrete
rotational symmetry of the lattice. Vortices enter the system as the
angular velocity is ramped up. The number of vortices is obtained 
from the phase winding of the condensate wavefunction
around the perimeter of the system. A lattice of size $L\times L$ can
contain at most $(L-1)$ vortices in the lowest band model.  We see
quantum phase transitions as the quasi-angular momenta of the ground
state changes. These are associated with vortices entering a system
which has filling incommensurate with the symmetry of the lattice.

Even though we have studied small quantum systems, our work has
implications in a broader context. The rotating lattice system is a
promising experimental approach allowing one to access more easily the
regime of strong quantum-correlations, a major goal in the field in
recent years. The novel aspect---the angular rotation frequency of an
optical lattice---provides an additional parameter which in
principle allows the experimentalist to explore a new axis of phase space.

In order to extend our calculations to study systems which are of the
size which will be more typical in experiments, it is necessary to go
beyond the exact quantum ground state calculations. Although this is
an important avenue for future work, the basic features of the quantum
phase transitions and the emergence of the vortex lattice should
persist in larger systems. Here, we have not explored in depth the
effects of high rotation which include the fragmentation of the
condensate and the emergence of the physics of the fractional quantum
Hall effect. In order to access this regime one should go beyond the
lowest band description.

\section{Acknowledgments}

We would like to thank John Cooper, Erich Mueller, Volker Schweikhard,
Shih-kuang Tung, Fei Zhou, and Peter Zoller for useful discussions. The authors
would like to acknowledge funding support from the Department of
Energy, Office of Basic Energy Sciences via the Chemical Sciences,
Geosciences, and Biosciences Division (R.B.), the National Science
Foundation (B.T.S., L.D.C.), NASA (B.M.P.), and Deutsche
Forschungsgemeinschaft (M.K.).


\begin{thebibliography}{10}
\bibitem{note:twostate} The lower/upper boundary of the $p$th Mott
zone is obtained by applying the two-state approximation to the
$p$th/$(p+1)$th band, where $p=1,2,\dots$ is the band index. The tip
of the zone, given by the intersection of the two boundaries, yields a
good estimate of the value of $t/U$ at which the superfluid-insulator
transition occurs at integer filling.
\end{thebibliography}
\end{document}